\documentstyle[preprint,aps,pra,epsbox]{revtex}
\begin{document}
\draft
\preprint{TKU-97-3 Feb. 1997}
\title{Spectroscopy of Heavy Mesons Expanded in $1/m_Q$}
\author{Takayuki Matsuki
\thanks{E-mail: matsuki@tokyo-kasei.ac.jp}}
\address{Tokyo Kasei University,
1-18-1 Kaga, Itabashi, Tokyo 173, JAPAN}
\author{Toshiyuki Morii
\thanks{E-mail: morii@kobe-u.ac.jp}}
\address{Faculty of Human Development,
Kobe University, 3-11 Tsurukabuto,\\ Nada, Kobe 657, JAPAN}
\maketitle
\begin{abstract}
Operating just once with the naive Foldy-Wouthuysen-Tani transformation on 
the relativistic Fermi-Yang equation for $Q\bar q$ bound states described by 
the semi-relativistic Hamiltonian which includes Coulomb-like as well as 
confining scalar potentials, we have calculated heavy meson mass spectra of $D$
and $B$ together with higher spin states. Based on the formulation recently 
proposed, their masses and wave functions are expanded up to the second order 
in $1/m_Q$ with a heavy quark mass $m_Q$ and the lowest order equation is 
examined carefully to obtain a complete set of eigenfunctions for the 
Schr\"odinger equation. Heavy quark effective theory parameters, 
$\bar\Lambda$, $\lambda_1$, and $\lambda_2$, are also determined at the first 
and second order in $1/m_Q$. 
\end{abstract}
\medskip
\section{Introduction}
Hadrons are composed of quarks and anti-quarks and are considered
to be governed by Quantum Chromodynamics, at least in principle. 
Since QCD describes a strong coupling interaction, perturbative
calculation of physical quantities of hadrons is not so reliable
other than the deep inelastic region where the coupling constant
becomes weak due to asymptotic freedom
and hence other methods like lattice gauge theory have been developed
to take into account nonperturbative effects.
However, the situation has dramatically changed when it is 
discovered that the system of heavy hadrons, composed of one heavy quark $Q$ 
and light quarks $q$ or anti-quarks $\bar q$, can be systematically expanded 
in $1/m_Q$ with a heavy quark mass, $m_Q$. The numerator of this expansion 
in $1/m_Q$ could be either $\Lambda_{\rm QCD}$ or $m_q$. 

This theory, HQET (Heavy Quark Effective Theory),\cite{HQET} 
is applied to many aspects of high energy theories and many kinds of physical
quantities of QCD which can be perturbatively calculated in $1/m_Q$.
Especially those regarding $B$ meson physics, e.g.,
the lowest order form factor ( which is now called Isgur-Wise function ) of
the semileptonic weak decay process $B \rightarrow D \ell\nu$ and
the Kobayashi-Maskawa matrix element $V_{cb}$, have been calculated
by many people.\cite{REVIEW}
However, since applications of HQET to higher order perturbative calculations
are very restricted, only forms of higher order operators are obtained.
Their Wilson coefficients are calculable but some of the matrix elements 
of those operators are obtained so that the whole quantity be somehow fitted 
with the experimental data.\cite{POL} This is because most of the calculations
based on HQET do not introduce realistic heavy meson wave functions and hence
there is no way to determine those quantities completely within the model.

In the former papers \cite{Matsuki1,Matsuki2}, using the Foldy-Wouthuysen-Tani
transformation \cite{FWT} we have developed a formulation
so that the Schr\"odinger equation for a $Q\bar q$ bound state 
can be expanded in terms of $1/m_Q$, i.e., 
the resulting eigenvalues as well as wave functions are
obtained order by order in $1/m_Q$.
In this paper, as one of the applications of our formulation 
we will calculate heavy meson spectrum of $D$ 
and $B$, and their higher spin states.
In order to do so, we would like to start from introducing 
phenomenological dynamics, i.e.,
assuming Coulomb-like vector and confining scalar potentials to 
$Q\bar q$ bound states (heavy mesons), expand a hamiltonian 
in $1/m_Q$ then perturbatively solve Schr\"odinger equation 
in $1/m_Q$. Angular part 
of the lowest order wave function is exactly solved.
After extracting asymptotic forms of the lowest order wave function
at both $r \rightarrow 0$ and $r \rightarrow \infty$
and adopting the variational method, we numerically obtain
radial part of the trial polynomial wave function
which is expanded in  powers of radial variable $r$.
Then fitting the smallest eigenvalues 
of a hamiltonian with masses of $D$ and $D^*$ mesons, a strong 
coupling $\alpha_s$ and other parameters
included in scalar and vector potentials are determined uniquely.
Using parameters obtained this way, other mass levels are calculated and
compared with the experimental data for $D/B$ mesons up to the second
order of perturbation. The lowest degenerate eigenvalues of the Schr\"odinger
equation gives the so-called $\bar \Lambda$ parameters for $u$, $d$ and $s$
light quarks, which is defined by
\[
  {\bar \Lambda}=\lim_{m_Q\to\infty}\left(E_H-m_Q\right)
\]
where $E_H$ is a calculated heavy meson mass and $m_Q$ a heavy quark 
mass.\cite{POL} Meson wave functions obtained thereby and expanded 
in $1/m_Q$ may be used to calculate ordinary form factors as well
as Isgur-Wise functions and its corrections in $1/m_Q$ 
for semileptonic weak decay processes. 

All the above calculations are calculated up to $1/m_Q^2$ and 
analyzed order by order in $1/m_Q$ to determine parameters as well as to
compare with results of Heavy Quark Effective Theory, e.g., the parameters,
$\lambda_1$, $\lambda_2$, and $\bar \Lambda$ in Sect.~\ref{sec:numeric}.
The final goal of this approach is to obtain higher order corrections to
Isgur-Wise functions, decay constants of heavy mesons, and the 
Kobayashi-Maskawa matrix element, $V_{cb}$, by using wave functions 
of heavy mesons obtained so that heavy meson spectrum is fitted with the
experimental data.

Below in Sects.~\ref{sec:intro} and \ref{sec:pert} we will first give 
formulation of this study and next in Sects.~\ref{sec:numeric} and
\ref{sec:com} give quantitative and qualitative discussions on the obtained
results.

\section{Hamiltonian}
\label{sec:intro}
The hamiltonian density for our problem is given by
\begin{eqnarray}
&{\cal H}_0=\int {dx^3}\,\left[\,q^{\dagger\,c}(x)\,\left(\vec{\alpha}_q
  \cdot\vec{p}_q+\beta_q m_q\right)\,q^c (x)+Q^\dagger (x)\left(\vec{\alpha}_Q
  \cdot\vec{p}_Q+\beta_Q m_Q\right) Q(x)\,\right],& \\
&{\cal H}_{\rm int}=\int\int {dx^3}{d{x^\prime}^3}\,
  {\bar {q^c}} (x)\,O_i\, q^c(x)\,V_i(x-x^\prime)\,
  {\bar Q} (x^\prime)\,O_i\,Q (x^\prime),&
\end{eqnarray}
where we consider only a scalar confining potential, $O_s=1, V_s=S(r)$, 
and a vector potential, $O_v=\gamma_\mu, V_v=V(r)$, with a relative radial 
variable $r$, which we think the best choice to phenomenologically describe 
the meson mass levels.\cite{Morii1,Morii2} The state of $Q\bar q$ is defined by
\begin{equation}
   \left| \psi\right> = \int d^3 x \int d^3 y~\psi_{\alpha \beta}(x-y)~
  {q_\alpha^c~}^\dagger (x)~Q_\beta^\dagger (y) 
  \left| 0 \right>,\label{eq:state}
\end{equation}
where $q^c(x)$ is a charge conjugate field of a light quark $q$ and 
the conjugate state of $Q\bar q$ by 
$\left< \psi\right|=\left| \psi\right>^\dagger$ 
with $\left<0\right|\equiv\left|0\right>^\dagger$.
From these definitions, we obtain the Schr\"odinger equation as
\begin{equation}
  H\, \psi=(m_Q+\tilde E)\,\psi,\label{eq:schrod1}
\end{equation}
where the bound state mass, $E$, is split into two parts, $m_Q$ and $\tilde E$
($=E-m_Q$), so that it expresses the fact that the heavy quark mass is
dominant in the bound state, $Q\bar q$, and $\psi$ is nothing but
the wave function which appears in the rhs of Eq.~(\ref{eq:state}).

\begin{figure}
\center\psbox[hscale=0.5,vscale=0.5]{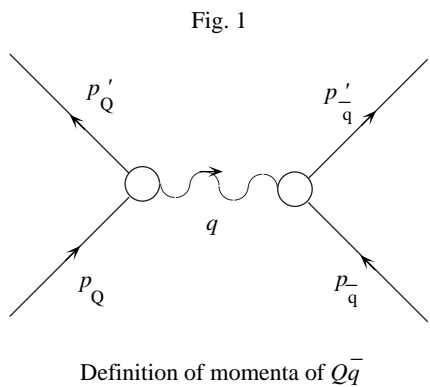}
\caption{Each momentum is defined.}
\label{Fig0}
\end{figure}

Operating with the FWT transformation and a charge conjugation operator,
which are defined in the Appendix \ref{app:schrodinger},
only on a heavy quark sector in this equation at the center of the 
mass system of a bound state,
one can modify the Schr\"odinger equation given by Eq.~(\ref{eq:schrod1}) as,
\begin{equation}
  \left(H_{FWT}-m_Q\right)~\otimes\psi _{FWT}
  =\tilde E\;\psi _{FWT},\label{eq:schrod2}
\end{equation}
where a notation $\otimes$ is introduced to denote that gamma matrices of
a light anti-quark is multiplied from left while those of a heavy quark
from right. The problem of this paper is to solve this equation, 
Eq.~(\ref{eq:schrod2}) in powers of $1/m_Q$.
As described first in this section, interaction terms are given by 
a confining scalar potential and a Coulomb vector potential with
transverse interaction\cite{Bethe} and a total hamiltonian is given by
\begin{eqnarray}
  H&&=\left( {\vec \alpha _q\cdot \vec p_q
  +\beta _qm_q} \right) +\left( 
  {\vec \alpha _Q\cdot \vec p_Q
  +\beta _Qm_Q} \right) +\beta _q\beta _Q\,S \nonumber \\
  &&\qquad +\left\{ {1-{1 \over 2}\left[\, {\vec \alpha _q\cdot
  \vec \alpha _Q+\left( {\vec \alpha _q\cdot \vec n} \right)\left( {\vec \alpha
  _Q\cdot \vec n} \right)}\, \right]} \right\}V, \label{eq:hamil}
\end{eqnarray}
where scalar and vector potentials are given by
\begin{equation}
  S\left( r \right)={r \over {a^2}}+b,\quad V\left( r \right)
  =-{4 \over 3}{\alpha _s \over r}
  \quad {\rm and}\quad\vec n={\vec r\over r}, \label{eq:svn}
\end{equation}
and the vector potential is averaged over longitudinal as well as transverse
as given in the last term of Eq.~(\ref{eq:hamil}).
The transformed hamiltonian is expanded in $1/m_Q$ as
\begin{equation}
  H_{FWT}-m_Q = H_{-1}+H_0+H_1+H_2+\cdots ,
\end{equation}
where
\begin{mathletters}
\label{eq:hameqs}
\begin{eqnarray}
  H_{-1}&=&-\left( {1+\beta _Q} \right)m_Q,\label{eq:ham_1}\\
  H_0&=&\vec \alpha _q\cdot \vec p+\beta _q m_q- 
  \beta _q\beta _Q\, S+\left\{ {1+{1 \over 2}\left[\, {\vec \alpha _q\cdot
  \vec \alpha _Q+\left( {\vec \alpha _q\cdot \vec n} \right)\left( {\vec \alpha
  _Q\cdot \vec n} \right)}\, \right]} \right\}V, \\
  H_1&=&-{1 \over {2\,m_Q}}\beta _Q\,{\vec p~}^2+
  {1\over m_Q}\beta _q\,\vec \alpha_Q\cdot 
  \left(\vec p+{1\over 2}\vec q\right)\, S
  +{1\over 2 m_Q}\vec\gamma_Q\cdot\vec q\,V \nonumber \\
  && -{1\over 2 m_Q} \left[\beta_Q \left(\vec p+{1\over 2}\vec q\right)+
  i\,\vec q\times\beta_Q\,\vec\Sigma_Q\right] \cdot\left[\,\vec\alpha_q+
  \left(\vec\alpha_q\cdot\vec n \right)\vec n\,\right] V, \\
  H_2&=& {1\over 2m_Q^2}\,\beta_q\beta_Q\,
  \left(\vec p+{1\over 2}\vec q\right)^2 S-{i\over 4m_Q^2}
  \vec q\times \vec p\cdot\beta_q\beta_Q\,\vec\Sigma_Q\, S-
  {1\over 8m_Q^2}{\vec q~}^2 V-{i\over 4m_Q^2}\vec q \times
  \vec p\cdot\vec\Sigma_Q\, V\nonumber \\
  && -{1\over 8m_Q^2}\left[\left(\vec p+\vec q\right)\,
  \left(\vec\alpha_Q\cdot\vec p\right)
  +\vec p\left(\vec\alpha_Q\cdot\left(\vec p+\vec q\right)\,\right)+
  i\, \vec q\times\vec p\,\gamma^5_Q\right]\cdot \left[\,\vec\alpha_q+
  \left(\vec\alpha_q\cdot\vec n\right)\vec n\,\right]\,V, \\
  \vdots\nonumber
\end{eqnarray}
\end{mathletters}
Here $H_i$ stands for the $i$-th order expanded hamiltonian and
since a bound state is at rest,
\begin{equation}
    \vec p=\vec p_q=-\vec p_Q,\quad 
    {\vec p}~^\prime ={\vec p_q}~^\prime=-{\vec p_Q}~^\prime,\quad
    \vec q={\vec p}~^\prime - \vec p,
\end{equation}
are defined, where primed quantities are final momenta and the relation
of these momenta with particles is depicted in Fig. \ref{Fig0}.

%
%

Details of derivation of equations in this section are given in the 
Appendix \ref{app:schrodinger}.

\section{Perturbation}
\label{sec:pert}
Using the hamiltonian obtained in the last section, we give in this
section the Schr\"odinger equation order by order in $1/m_Q$.
Details of the derivation in this section are given in 
the Appendix \ref{app:pert}. First we introduce projection
operators:
\begin{equation}
    \Lambda_\pm = {{1\pm \beta_Q}\over 2},
\end{equation}
which correspond to positive-/negative-energy projection operators
for a heavy quark sector at the rest frame of a bound state. These are
given by $(1\pm v{\kern-6pt /})/2$ in the moving frame of a bound state
with $v^\mu$ the four-velocity of a bound state.
Then we expand the mass and wave function of a bound state in $1/m_Q$ as
\begin{eqnarray}
  \tilde E=E-m_Q&=&E^\ell_0+E^\ell_1+E^\ell_2+\ldots,\label{eq:mass}\\
  \psi_{FWT}&=&\psi^\ell_0+\psi^\ell_1+\psi^\ell_2+\ldots,\label{eq:wf}
\end{eqnarray}
where $\ell$ stands for a set of quantum numbers that distinguish independent
eigenfunctions of the lowest order Schr\"odinger equation, and a
subscript $i$ of $E^\ell_i$ and $\psi^\ell_i$ stands for the order of $1/m_Q$.
\subsection{-1st order}
\label{subsec:-1st}
The -1st order Schr\"odinger equation in $1/m_Q$ gives
\begin{equation}
    \psi^\ell_0=\Lambda_-\otimes\psi^\ell_0,\label{eq:-1st}
\end{equation}
whose explicit form is solved in the Appendix \ref{app:0thsol} and is given by
\begin{equation}
  \psi _0^\ell=\Psi_\ell^+=\left(\matrix{0&\Psi _{j\,m}^k
  (\vec r)}\right).\label{eq:psi}
\end{equation}
Here $\ell$ stands for a set of quantum numbers, $j$, $m$, and $k$ and
\begin{equation}
  \Psi _{j\,m}^k(\vec r) ={1\over r}\left( {\matrix{u_k(r)\cr 
  -i\,v_k(r)\left(\vec \sigma\cdot\vec n\right)\cr}} \right)
  \;y_{j\,m}^k (\Omega),\label{eq:Psi+}
\end{equation}
where $j$ is a total angular momentum of a meson, $m$ is its $z$ component,
$k$ is a quantum number which takes only values, $k=\pm j,\;\pm(j+1)\;
{\rm and}\;\ne 0$, and $u_k(r)$ and $v_k(r)$ are polynomials of a radial 
variable $r$. $y_{j\,m}^k (\Omega)$ are functions of angles and spinors 
of a total angular momentum, $\vec j=\vec l+\vec s_q+\vec s_Q$.
The corresponding operator for the quantum number $k$ is given by
\begin{equation}
  -\beta_q\left(\vec\Sigma_q\cdot\vec\ell+1\right),\label{eq:ok}
\end{equation}
which satisfies
\begin{equation}
  -\beta_q\left(\vec\Sigma_q\cdot\vec\ell+1\right)
  \left(\matrix{0&\Psi _{j\,m}^k(\vec r)}\right) = k
  \left(\matrix{0&\Psi _{j\,m}^k(\vec r)}\right),\label{eq:k1}
\end{equation}
i.e.,
\begin{equation}
  \left[-\beta_q\left(\vec\Sigma_q\cdot\vec\ell+1\right),\,H_0^{--}
  \right]=0,\label{eq:k2}
\end{equation}
with $H_0^{--}$ being given in the Appendix \ref{app:matrix},
the lowest order non-trivial hamiltonian,
\[
  H_0^{--}\otimes\psi _0^\ell=E_0^k\psi _0^\ell.
\]

Note that since charge conjugation operates on the heavy quark
sector the $\Lambda_-$ projection operator appears in 
Eq.\ (\ref{eq:-1st}), i.e., positive components of $Q$ corresponds
to negative components of $U_c\,Q$.
\subsection{Zero-th order}
\label{subsec:0-th}
The zero-th order equations are given by
\begin{eqnarray}
  &\left[\,\vec\alpha_q\cdot\vec p+\beta_q\left(m_q+S\right)+V\,\right]
  \otimes\psi _0^\ell=E_0^\ell\,\psi _0^\ell,&\label{0th:1}\\
  &-2m_Q\Lambda _+\otimes\psi _1^\ell
  +{1 \over 2}\Lambda_-\left[\, {\vec \alpha _q\cdot\vec \alpha _Q
  +\left( {\vec \alpha _q\cdot \vec n} \right)\left( {\vec \alpha
  _Q\cdot \vec n} \right)}\, \right] V\otimes\psi _0^\ell
  =0.&\label{0th:2}
\end{eqnarray}
Eq.\ (\ref{0th:1}) gives the lowest non-trivial Schr\"odinger equation
with a solution given by Eq.\ (\ref{eq:psi}) and $\vec n=\vec r/r$ . 
Detailed analysis of this equation is given in 
Appendix \ref{app:0thsol}. $\Lambda_+$ components of wave functions can 
be expanded in terms of the eigenfunctions,
\begin{equation}
  \Psi_\ell^-=\left(\matrix{\Psi _{j\,m}^k(\vec r)&0}\right).
  \label{eq:Psi-}
\end{equation}
Expanding $\Lambda_+\otimes\psi _1^\ell$ in terms of 
this set of eigenfunctions, one can obtain the solution for 
Eq.~(\ref{0th:2}) as
\begin{equation}
    \Lambda _+\otimes\psi _1^\ell=\sum\limits_{\ell\,'}
    {c_{1-}^{\ell \,\ell\,'} \Psi _{\ell\,'}^-},
\end{equation}
with the coefficients,
\begin{equation}
  c_{1-}^{\ell \,\ell\,'}={1\over 4m_Q}\left<{\Psi^-_{\ell\,'}}\right|
  \left[\, {\vec \alpha _q\cdot\vec \alpha _Q+\left( 
  {\vec \alpha _q\cdot \vec n} \right)\left( {\vec \alpha_Q\cdot \vec n} 
  \right)}\, \right]V\left|\Psi^+_\ell\right>.\label{eq:c1-}
\end{equation}
Here the inner product is defined to be
\begin{equation}
    \left<{\Psi ^\alpha_\ell}\right|O\left|\Psi^\beta_{\ell\,'}\right>
    =\int d^3r\, {\rm tr}\left({\Psi^\alpha_\ell}^\dagger\left(
    O\otimes\Psi^\beta_{\ell\,'}\right)\right),
\end{equation}
and the zero-th order wave functions are normalized to be 1, 
\begin{equation}
     \left<{\Psi ^\alpha_\ell}\right.\left|\Psi^\beta_{\ell\,'}\right>
     =\delta_{\ell\,{\ell\,'}}\delta^{\alpha\,\beta}
     \quad {\rm for}\quad\alpha,\,\beta=+\;{\rm or}\,-.\label{nomal0}
\end{equation}
\subsection{1st order}
\label{subsec:1st}
The 1st order equation is given by
\begin{equation}
    -2m_Q\Lambda _+\otimes\psi _2^\ell+H_0\otimes\psi _1^\ell+
    H_1\otimes\psi _0^\ell =E_0^\ell\psi _1^\ell+
    E_1^\ell\psi _0^\ell. \label{eq:1st:1}
\end{equation}
Multiplying projection operators $\Lambda_\pm$ from right with the
above equation, and expanding 
$\psi_1^\ell$ in terms of $\Psi_\ell^\pm$ as
\begin{equation}
  \psi_1^\ell = \sum\limits_\ell\,\left(c_{1+}^{\ell \,\ell\,'}
  \Psi _{\ell\,'}^+ + c_{1-}^{\ell \,\ell\,'}\Psi _{\ell\,'} ^- \right),
\end{equation}
one obtains
\begin{equation}
  E_1^\ell=\sum\limits_{\ell\,'}\,c_{1-}^{\ell \,{\ell\,'}}
  \left<\Psi _\ell ^+ \right|\Lambda _+H_0\Lambda _-\left|
  \Psi _{\ell\,'} ^-\right>+\left<{\Psi _\ell ^+}\right| \Lambda _-H_1 
  \Lambda _-\left|\Psi _\ell ^+\right>, \label{eq:1st:energy}
\end{equation}
which gives the first order perturbation correction to the mass when
one calculates matrix elements of the rhs among eigenfunctions and
\begin{eqnarray}
  c_{1+}^{\ell \,k}&=&{1\over E_0^\ell-E_0^k}\left[
  \sum\limits_{\ell\,'}\,c_{1-}^{\ell \,{\ell\,'}}
  \left<\Psi _k ^+ \right|\Lambda _+H_0\Lambda _-\left|
  \Psi _{\ell\,'} ^-\right>+  \left<{\Psi _k ^+}\right| \Lambda _-H_1 
  \Lambda _-\left|\Psi _\ell ^+\right>\right],
  \quad {\rm for}~k\ne\ell \label{eq:c1+kl} \\
  c_{1+}^{k\,k}&=&0.\label{eq:c1+kk}
\end{eqnarray}
This completes the solution for $\psi _1^\ell$ since $\Lambda_-\psi^\ell_1$,
or $c_{1-}^{\ell\,{\ell\,'}}$, is obtained in the last subsection.
Here we have used the normalization for the total wave function, 
$\psi^\ell$, as
\begin{equation}
  \left<\psi^\ell\right|\left.\psi^{\ell\,'}\right>
  =\delta_{\ell\,{\ell\,'}},\label{eq:normpsi}
\end{equation}
where we have neglected color indices in this paper and hence a color 
factor, $N_c=3$, in the above equation since it does not change the essential
arguments. This definition of Eq.~(\ref{eq:normpsi}) is admitted 
because here we are not calculating the absolute value of 
the form factors. The appropriate normalization (normally given by
$2E$ with a bound state mass $E$) will be adopted
in future papers in which we will calculate some form factors.
This way of solving Eq.\ (\ref{eq:1st:1}) is unique and we will 
use this method below to solve similar equations appearing subsection
Actually this method has been already used to obtain 
Eqs.~(\ref{0th:1}, \ref{0th:2}) and to solve Eq.~(\ref{0th:2}) obtaining
the coefficients $c_{1-}^{\ell\,\ell\,'}$ by Eq.~(\ref{eq:c1-}).

One obtains $\Lambda_+\otimes\psi_2^\ell$ as in the former subsection,
\begin{equation}
    \Lambda _+\otimes\psi _2^\ell=\sum\limits_{\ell\,'}
    {c_{2-}^{\ell \,{\ell\,'}}\Psi _{\ell\,'} ^-},
\end{equation}
with the coefficients,
\begin{equation}
    c_{2-}^{\ell \,{\ell\,'}}={1 \over {2m_Q}}
    \left<{\Psi^- _{\ell\,'}}\right.\left|
    \left(\left(H_0-E^\ell_0\right)\Lambda _+\otimes\psi^\ell_1 
    +H_1\Lambda _+\otimes\psi^\ell_0\right)\right>.\label{eq:c2-}
\end{equation}
\subsection{2nd order}
\label{subsec:2nd}
The 2nd order equation is given by
\begin{equation}
    -2m_Q\Lambda _+\otimes\psi _3^\ell+H_0\otimes\psi _2^\ell+
    H_1\otimes\psi _1^\ell+H_2\otimes\psi _0^\ell
    =E_0^\ell\psi _2^\ell+E_1^\ell\psi _1^\ell+E_2^\ell\psi_0^\ell.
\end{equation}
As in the above case (1st order), we obtain
\begin{equation}
  E_2^\ell=
  \sum\limits_{\ell\,'}\,c_{2-}^{\ell \,{\ell\,'}}\left<
  {\Psi _\ell ^+}\right| \Lambda _+H_0 \Lambda _-\left|
  \Psi _{\ell\,'} ^-\right>+\left< {\Psi _\ell ^+}\right| H_1\Lambda _- 
  \left.\otimes\psi _1 ^\ell\right>+\left<{\Psi _\ell ^+}\right| \Lambda _
  -H_2 \Lambda _- \left|\Psi _\ell ^+\right>,\label{eq:2nd:energy}
\end{equation} 
which gives the second order perturbation corrections to the mass and
\begin{eqnarray}
  c_{2+}^{\ell\,k}&=&{1\over E_0^\ell-E_0^k}\left[
  \sum\limits_{\ell\,'}\,c_{2-}^{\ell \,{\ell\,'}}\left<
  \Psi _k ^+\right| \Lambda _+H_0 \Lambda _-\left|\Psi _{\ell\,'} ^-\right>+
  \left< \Psi _k ^+\right| \Lambda _+H_1 \left.\otimes\psi _1 ^\ell\right>
  \right. \nonumber \\
  &&\left.+ 
  \left<\Psi _k ^+\right| \Lambda _-H_2\Lambda _-\left|\Psi _\ell ^+\right>
  -E_1^\ell\,c_{1+}^{\ell\,k}\right], \qquad {\rm for}~k
  \ne\ell\label{eq:c2+kl}\\
  c_{2+}^{k\,k}&=&
  -{1\over 2}\sum\limits_\ell\left(\left| c_{1+}^{k\,\ell}\right|^2
  +\left| c_{1-}^{k\,\ell}\right|^2 \right).\label{eq:c2+kk}
\end{eqnarray}
This completes the solution for $\psi _2^k$ since $\Lambda_-\psi^k_2$,
or $c_{2-}^{\ell\,{\ell\,'}}$, is obtained in the last subsection.

Although we do not need in this paper, one obtains 
$\Lambda_+\otimes\psi_3^\ell$ as
\begin{equation}
    \Lambda _+\otimes\psi _3^\ell=\sum\limits_{\ell\,'}
    {c_{3-}^{\ell \,{\ell\,'}}\Psi _{\ell\,'}^-},
\end{equation}
with the coefficients,
\begin{equation}
    c_{3-}^{\ell \,{\ell\,'}}={1 \over {2m_Q}}
    \left<{\Psi^-_{\ell\,'}}\right.\left|
    \left(\left(H_0-E^\ell_0\right)\Lambda _+\otimes\psi^\ell_2
    +\left(H_1-E^\ell_1\right)\Lambda_+\otimes\psi^\ell_1
    +H_2\Lambda _+\otimes\psi^\ell_0 \right)\right>.\label{eq:c3-}
\end{equation}
\section{Numerical Analysis}
\label{sec:numeric}
In this section, we give numerical analysis of our analytical calculations
obtained in the former sections order by order in $1/m_Q$.
In order to solve Eq.\ (\ref{0th:1}), we have to numerically obtain
a radial part of the wave function, 
$\Psi_\ell^+=\left(\matrix{0&\Psi_{j\,m}^k}\right)$, given by
\begin{equation}
  \Psi _{j\,m}^k(\vec r) ={1\over r}\left( {\matrix{u_k(r)\cr 
  -i\,v_k(r)\left(\vec \sigma\cdot\vec n\right)\cr}} \right)
  \;y_{j\,m}^k (\Omega),
\end{equation}
detailed properties of which are described in the Appendix \ref{app:0thsol}.
As described in the same Appendix, the lowest order, non-trivial Schr\"odinger
equation is reduced into Eq.\ (\ref{eq:app:schrod3}),
\begin{equation}
  \left( {\matrix{{m_q+S+V}&{-\partial _r+{k \over r}}\cr
  {\partial _r+{k \over r}}&{-m_q-S+V}\cr}} \right)
  \left( {\matrix{{u_k\left( r \right)}\cr{v_k\left( r \right)}\cr}} 
  \right)=E^k_0\left( {\matrix{{u_k\left( r \right)}\cr
  {v_k\left( r \right)}\cr}} \right),\label{eq:schrod3}
\end{equation}
This eigenvalue equation is numerically solved by taking into 
account the asymptotic behaviors at both $r \rightarrow 0$ and
$r\rightarrow \infty$ and the forms of $u_k(r)$ and $v_k(r)$ are given by
\begin{equation}
  u_k(r),\; v_k(r) \sim w_k(r)\, \left({r\over a}\right)^\gamma\,\exp\left(
  -\left(m_q+b\right)\,r-{1 \over 2}\left({r\over a}\right)^2 \right),
  \label{eq:waver}
\end{equation}
where
\begin{equation}
  \gamma=\sqrt{k^2-\left({4\alpha_s \over 3}\right)^2}
\end{equation}
and $w_k(r)$ is a finite series of a polynomial of $r$
\begin{equation}
  w_k(r)=\sum_{i=0}^{N-1}\;a^k_i\,\left({r\over a}\right)^i,\label{eq:wk}
\end{equation}
which takes different coefficients for $u_k(r)$ and $v_k(r)$.

i)~We have fixed the value of a light quark mass, $m_q$, to be $0.01$
GeV as listed in Table \ref{table1} since only in the vicinity of this value
the $D$ and $D^*$ masses can be fitted with the experimental
values. We believe that when the mass, $m_q=m_u=m_d$, is running with
momentum these are the correct current quark mass since the momentum
is given by the order of the $B$ meson mass ($\sim 5$ GeV)\cite{Politzer}
though we have not used the running mass to solve the Schr\"odinger equation.
We have adopted $N-1=7$ for the highest power of $r$ that gives
sixteen solutions to Eq.~(\ref{eq:schrod3}), half of which corresponds
to negative energies and another half to positive ones of $q^c$ state. 
The lowest eigenvalue of the positive energies is assigned to the 
physical state. That is, although 
we have a node quantum number, $n$, other than $j$, $m$, and $k$, for 
Eq.~(\ref{eq:schrod3}), we take only the $n=0$ solution for each value 
of $k$ and $j$ quantum numbers and we do not take into account higher 
order node solutions in this paper. The lowest positive eigenvalue 
solution for $N=8$ gives the one closest to the zero node compared 
with other $N$. 
That is, other $N$ gives a rather oscillatory solution.

In the case of a Hydrogen atom, for instance, only the Coulomb
potential $V\sim 1/r$ survives in the above problem and 
a radial function, $w_k(r)$, becomes a hypergeometric function and
its finite series of a polynomial gives discrete energy levels.
In our case, since the potential includes a scalar term we can not 
analytically solve the above reduced Schr\"odinger equation, 
Eq.\ (\ref{eq:schrod3}). If we force to make 
the functions, $u_k(r)$ and $v_k(r)$, finite series and relate 
the coefficients of those functions via recursive equations, it leads us
to inconsistency among coefficients of each term, $r^i$, of a polynomial.
We just assume in this paper that $u_k(r)$ and $v_k(r)$ are trial finite 
polynomial functions of $r$.

ii)~To first determine the parameters, $\alpha_s$, $a$, and $b$ appearing 
in the potentials given by Eq.~(\ref{eq:svn}), and $m_c$, we have 
calculated the chi square defined by
\begin{equation}
  \chi^2={(M_D-E_D)^2\over \sigma_D^2}+
  {(M_{D^*}-E_{D^*})^2\over \sigma_{D^*}^2}, \label{eq:chi2}
\end{equation}
where $M_{D,\,D^*}$ and $E_{D,\,D^*}$ are the observed and calculated
masses of $D$ and $D^*$, respectively and $\sigma_{D,\,D^*}$ are the 
experimental errors for each meson mass. As mentioned already $m_q=m_d=m_u$
is fixed to be 10 MeV. Masses, $M_{D,\,D^*}$, are averaged over charges 
since we have not taken into account the electromagnetic interaction.
We have determined the values for these parameters which give the least
value of $\chi^2$, which is given in Table \ref{table2}.

iii)~The $s$ quark mass, $m_s$, is determined so that
the similar equation to Eq.~(\ref{eq:chi2}) takes the minimum value
where we substitute $D_s$ and $D_s^*$ meson masses instead of $D$ and $D^*$,
while the $b$ quark mass, 
$m_b$, is determined by $B$ and $B^*$ meson masses. The input values
to determine these parameters are given in Table \ref{table1}.
\begin{center}
  \fbox{Table I}
\end{center}

iv)~There are two types of solutions to optimal values for these parameters, 
i.e., one set for $b < 0$ which is listed in Table \ref{table2}, and another 
for $b > 0$. However, the solution for $b > 0$ gives large difference between
calculated values and observed ones for higher order spins and also gives
negative values for some spectrum even though the lowest lying sates are in
good agreement with the observed ones. Hence we disregard this set of 
parameters.

\begin{figure}
\center\psbox[hscale=0.5,vscale=0.5]{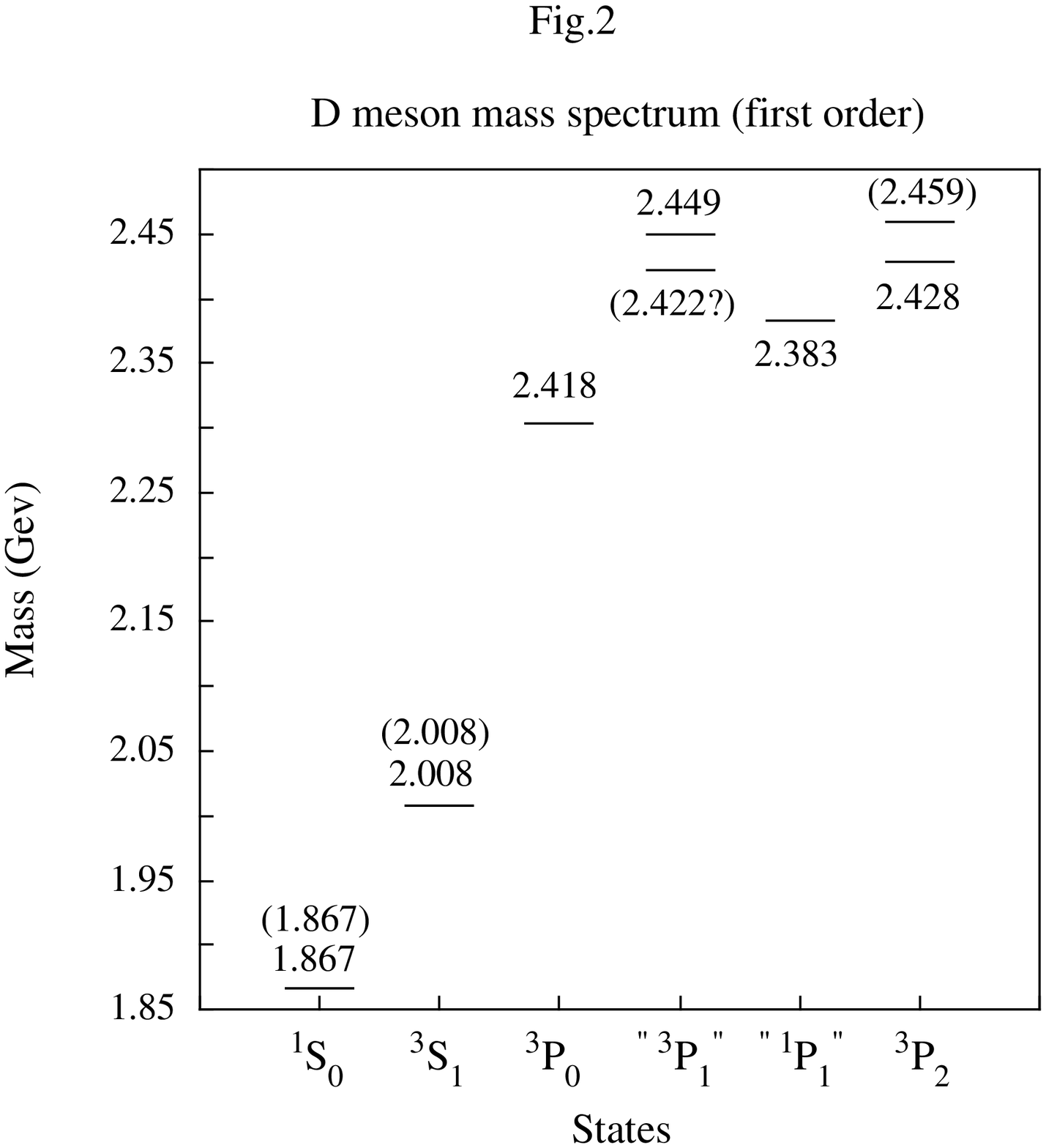}
\caption{The first order plot of $D$ meson masses. Values in the brackets are
the observed values.}
\label{Fig1}
\end{figure}

Tables \ref{table3} $\sim$ \ref{table10} give calculated values, 
$M_{\rm calc}$, together with the zero-th order masses, $M_0$ that are 
degenerate with the same value of $k$, ratios, $p_i/M_0$ and 
$n_i/M_0$, and the observed values, $M_{\rm obs}$. Here the heavy meson mass,
$E_H$, is expanded in $1/m_Q$ up to the $n$-th order as
\begin{equation}
  E_H=M_0+\sum_{i=1}^n\,p_i+\sum_{i=1}^n\,n_i,
\end{equation}
with $M_0=m_Q+E_0$ being the degenerate mass, $p_i$ the $i$-th order 
correction from positive components of a heavy meson wave function, and $n_i$
the $i$-th order correction from negative components. Note also that
the exponential factor in the brackets in the first row of each table 
should be multiplied with a value of each column except for those with the
explicit exponential factor.
\begin{center}
  \fbox{Table III $\sim$ X}
\end{center}
Strictly speaking each state is classified by two 
quantum numbers, $k$ and $j$, and also approximately classified by the upper
component of the light anti-quark sector in terms of the usual notation, 
$~^{2S+1}L_J$. Studying the functions $y_{j\,m}^k$ carefully, one finds the 
upper component of $\Psi_{j\,m}^k(\vec r)$ corresponds to the following 
Table \ref{table0}, respectively.
\begin{center}
  \fbox{Table XI}
\end{center}
Here $J$ in $J^P$ and $^{2S+1}L_J$ is the same as a total angular
momentum, $j$, in the Table \ref{table0}. 
Although the states can be completely classified in terms of two quantum 
numbers, $k$ and $j$, we would like ordinarily to classify them in terms of 
$^{2S+1}L_J$. However, the states classified by $J^P=1^+$ and $2^-$, are 
mixtures of two states in terms of $^{2S+1}L_J$ as given by the 
Table \ref{table0}. We would approximately regard the state ($k$, $j$)=(1, 1)
with $"^3P_1"$, \,(-2, 1) with $"^1P_1"$ and (2, 2) with $"^3D_2"$, 
respectively, whose legitimacy can be supported by calculating the coefficient
of each state $^{2S+1}L_J$ included in the mixture state. We denote 
them with double quotations so that they remind us an approximate 
representation of the state in terms of $^{2S+1}L_J$. Using 
Eq.~(\ref{eq:app:u}) in Appendix \ref{app:0thsol}, their relations are
given by
\begin{eqnarray}
  \left(\matrix{\left|"^3P_1"\right>\cr \left|"^1P_1"\right>\cr}\right)&=&
  {1\over \sqrt{3}} \left(\matrix{\sqrt{2}&1\cr -1&\sqrt{2}\cr}\right)
  \left(\matrix{\left|~^3P_1~\right>\cr \left|~^1P_1~\right>\cr}\right), \\
~\nonumber \\
  \left(\matrix{\left|"^3D_2"\right>\cr \left|"^1D_2"\right>\cr}\right)&=&
  {1\over \sqrt{5}} \left(\matrix{\sqrt{3}&\sqrt{2}\cr 
  -\sqrt{2}&\sqrt{3}\cr}\right)
  \left(\matrix{\left|~^3D_2~\right>\cr \left|~^1D_2~\right>\cr}\right).
\end{eqnarray}

v)~From Tables \ref{table3} $\sim$ \ref{table10} we see that the perturbative
calculation with these parameters might not work well for higher $k$. 
Namely masses of $1^-$ and $2^-$ for $k=+2$ give some odd values. 
They become even negative for 
$"^3D_2"$ of $D_s$ and $B_s$ at the second order as shown in Tables 
\ref{table8} and \ref{table10}. Hence we disregard in this paper all 
calculated masses of $^3D_1(1^-)$ and $"^3D_2"(2^-)$ states in any order. 
In order to remedy this problem, we may need to improve the potential form 
or adopt some other methods.\cite{Matsuki3} Hence here only the first order 
mass spectra are depicted without higher spin states ($^3D_1(1^-)$ and 
$"^3D_2"(2^-)$) in Figs.~\ref{Fig1}$\sim$\ref{Fig4}.
%
%
\begin{figure}
\center\psbox[hscale=0.5,vscale=0.5]{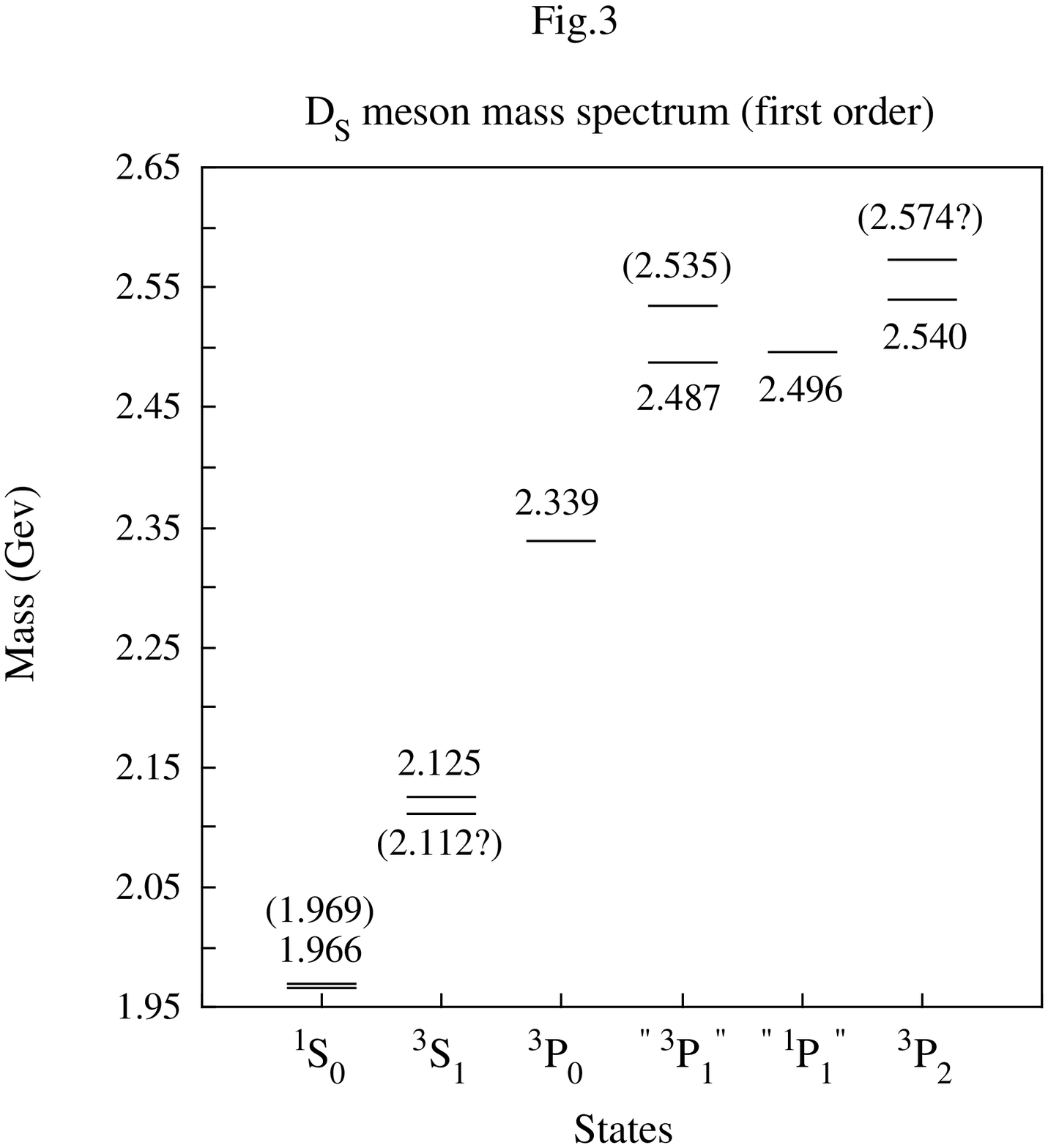}
\caption{The first order plot of $D_s$ meson masses. Values in the brackets 
are the observed values.}
\label{Fig2}
\end{figure}
\begin{figure}
\center\psbox[hscale=0.5,vscale=0.5]{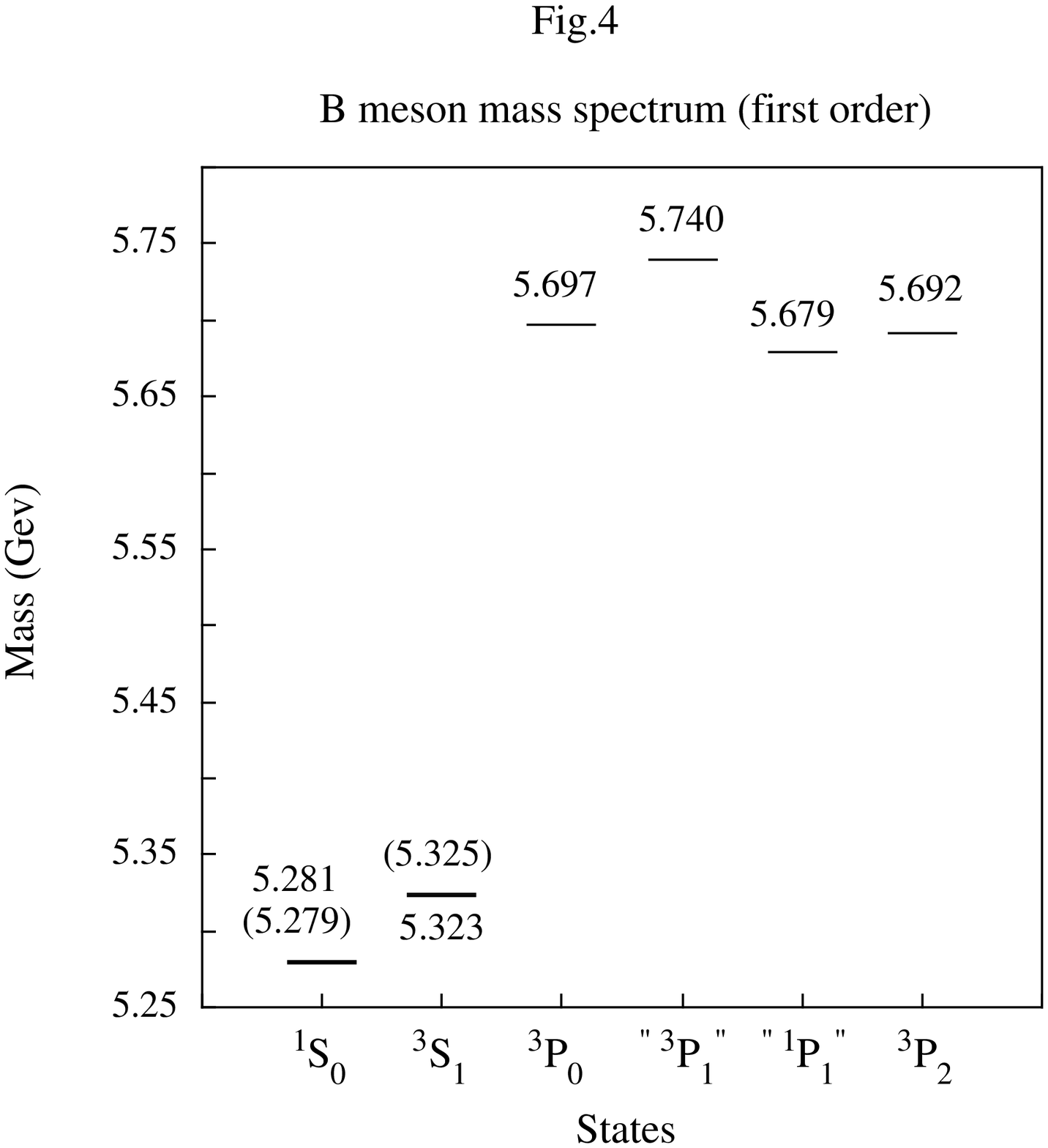}
\caption{The first order plot of $B$ meson masses. Values in the brackets 
are the observed values.}
\label{Fig3}
\end{figure}
\begin{figure}
\center\psbox[hscale=0.5,vscale=0.5]{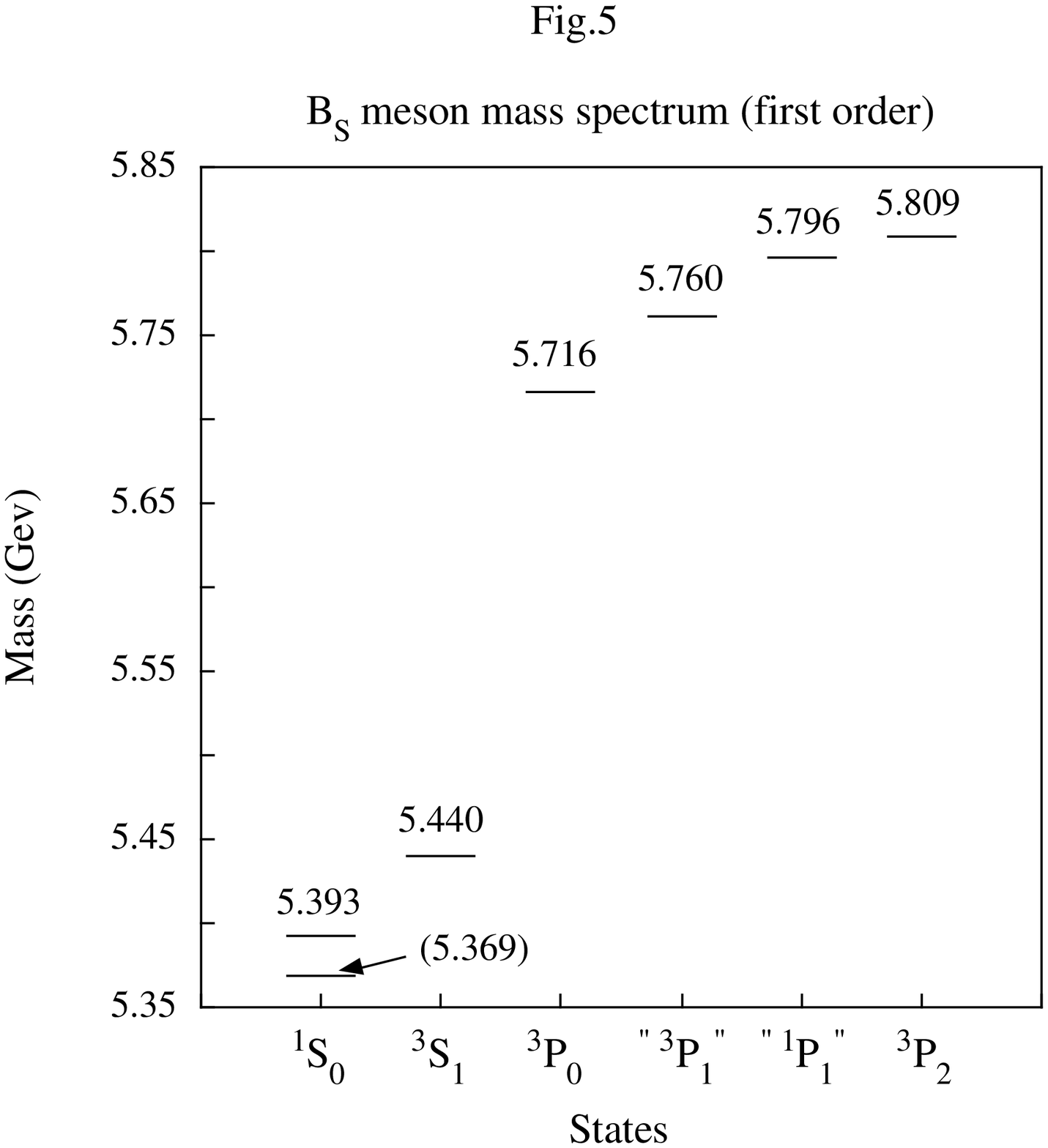}
\caption{The first order plot of $B_s$ meson masses. Values in the brackets 
are the observed values.}
\label{Fig4}
\end{figure}

One may also notice that the $s$ quark mass listed in Table \ref{table2}
is relatively small ($\sim$ 90 MeV) compared with the conventionally
used values, $\sim 150$ GeV, which is regarded as the current quark mass. 
It is interesting to note that these values are also obtained as $\bar m_s
(\mu=2~{\rm GeV})$ in the recent lattice QCD calculations.\cite{Lattice}

vi)~Two states, pseudoscalar ($0^-$) and vector ($1^-$), are degenerate
at the zero-th order in $1/m_Q$ since the eigenvalue $E_0^k$ for these
states depends on the same quantum number $k=-1$, which are split into two 
via the heavy quark spin interaction terms, like $-V'\left( {\vec \alpha _q
\cdot \vec \Sigma _Q\times \vec n}  \right)$ in $H_1^{--}$ and all 
terms in $H_1^{-+}$ given by Eq.~(\ref{eq:app:ham+-}). Similar resolution
of the degeneracy among the states with the same value of $k$ occurs via 
the same interaction terms.

vii)~The simple-minded heavy meson mass formula given by
\begin{equation}
  m_c\,\left(E_{D^*}-E_D\right) = m_b\,\left(E_{B^*}-E_B\right)
  \label{eq:mform}
\end{equation}
holds exactly at the first order calculation. This is because the zero-th
order mass of two states with the same $k$ is degenerate and by definition
the first order correction to this mass is proportional to $1/m_Q$
as given by $H_1$ of Eq.~(\ref{eq:hameqs}) or by $H_1^{\alpha\,\beta}$
of Eq.~(\ref{eq:app:ham+-}). To see Eq.~(\ref{eq:mform}) as an prediction, 
replacing $E_X$ with the observed values $M_X$ we obtain at the first order,
\begin{equation}
  {M_{B^*}-M_B \over M_{D^*}-M_D}={m_c \over m_b}=0.299,
\end{equation}
which should be compared with the experimental value, 0.326. This discrepancy
between the calculated and the observed comes from our calculation of
$B$ meson mass spectrum listed in Table~\ref{table5} which give $B$ and $B^*$
meson masses slightly different values from the observed ones.
Hold also similar equations to 
Eq.~(\ref{eq:mform}) for higher spin states with the same $k$ quantum number 
because of the same reason given above.

viii)~The so-called $\bar \Lambda$ parameter can be calculated using the 
definition\cite{POL}
\begin{equation}
  {\bar \Lambda}=\lim_{m_Q\to\infty}\left(E_H-m_Q\right)
  =M_0-m_Q=E_0^{-1} \label{Lambda}
\end{equation}
where $E_H$, $M_0$, and $m_Q$ are calculated heavy meson mass, the lowest 
degenerate bound state mass, and a heavy quark mass, respectively. Difference
of $M_0$ and $m_Q$ is nothing but the lowest leading eigenvalue, 
$E_0^k$, with $k=-1$ in our model. From Tables \ref{table3} and \ref{table4}
and $m_c=1.457$ listed in Table \ref{table2}, one obtains at the first order
\begin{eqnarray}
  {\bar \Lambda_{u,d}}&=&{M_0}_D-m_c={M_0}_{D^*}-m_c=0.412~ {\rm GeV},\\
  {\bar \Lambda_s}&=&{M_0}_{D_s}-m_c={M_0}_{D_s^*}-m_c=0.529~ {\rm GeV},
\end{eqnarray}
and from Tables \ref{table3} and \ref{table4} and $m_c=1.347$ listed in 
Table \ref{table2}, one obtains at the second order
\begin{eqnarray}
  {\bar \Lambda_{u,d}}&=&{M_0}_D-m_c={M_0}_{D^*}-m_c=0.518~ {\rm GeV},\\
  {\bar \Lambda_s}&=&{M_0}_{D_s}-m_c={M_0}_{D_s^*}-m_c=0.629~ {\rm GeV},
\end{eqnarray}
where ${M_0}_{D_{(s)}},\,{M_0}_{D_{(s)}^*}$ are the calculated lowest order 
$D$ meson mass defined by Eq.~(\ref{Lambda}).

ix)~Parameters which give nonperturbative corrections to inclusive
semileptonic $B$ decays are defined as \cite{INC1,INC2},
\begin{eqnarray}
  \lambda_1&=&{1\over 2m_Q}\left<H(v)\right|{\bar h_v}(iD)^2h_v
  \left|H(v)\right>,\\
  \lambda_2&=&{1\over 2d_H\,m_Q}\left<H(v)\right|{\bar h_v}{g\over 2}
  \sigma_{\mu\,\nu}G^{\mu\,\nu}h_v\left|H(v)\right>,
\end{eqnarray}
where $h_v$ is the heavy quark field in the HQET with velocity $v$.
$d_H\,=\,3,\,-1$ for pseudoscalar or vector mesons, respectively. Then
the heavy meson mass can be expanded in terms of the heavy quark mass,
$\lambda_1$, $\lambda_2$, and $\bar \Lambda$, as
\begin{equation}
  E_H=m_Q+{\bar \Lambda}-{\lambda_1+d_H\lambda_2\over 2m_Q}+\ldots.
\end{equation}
The first order calculation in $1/m_Q$ makes $2\lambda_2$ equal to 
Eq.~(\ref{eq:mform}) and the $\lambda_1$ can be calculated using the above
equation as
\begin{eqnarray}
  \lambda_1&=&2m_b\left(m_b+{\bar \Lambda_u}-{\tilde E_B}\right)
  -3\lambda_2, \label{eq:lambda1} \\
  \lambda_2&=&{m_b\over 2}\,\left({\tilde E_{B^*}}-{\tilde E_B}\right)
  \label{eq:lambda2} ,
\end{eqnarray}
where $\tilde E_B$ and  $\tilde E_{B^*}$ are the calculated $B$ meson masses 
without the second order corrections. The results are given by, at the 
first order, 
\begin{equation}
  \lambda_1=-0.378~{\rm GeV}^2, \qquad
  \lambda_2=0.112~{\rm GeV}^2.
\end{equation}
and at the second order,
\begin{equation}
  \lambda_1=-0.238~{\rm GeV}^2, \qquad
  \lambda_2=0.0255~{\rm GeV}^2.
\end{equation}
Here we notice that although the first term in Eq.~(\ref{eq:lambda1}) is 
expected to be $\sim O(1)$ we find it to be small from Table \ref{table5}
\underline{at the first order} and obtain the approximate relation, 
\begin{equation}
  \lambda_1\sim -3\lambda_2.
\end{equation}
These values should be compared with those in Ref.~\cite{INC3}, which
give $\bar\Lambda_u=0.39\pm0.11$ GeV, $\lambda_1=-0.19\pm0.10$ GeV$^2$, and
$\lambda_2\simeq 0.12$ GeV$^2$.

x)~Recently, it has been pointed out that the kinetic 
energy of heavy quark inside a heavy meson plays an important role in the 
determination of the ratios $f_B/f_D$, 
$\left(M_{B^*}-M_B\right)/\left(M_{D^*}-M_D\right)$, and
$\left| V_{ub}/V_{cb}\right|$, in which use has been made the Gaussian form 
for the heavy meson wave function and has been adopted the so-called Cornell
potential, the same as ours.\cite{Kim} They have derived the relation of
these physical quantities in terms of the Fermi momentum, $p_F$, introduced 
in \cite{Alt} in which $p_F$ is related to a heavy quark recoil momentum, 
$\vec p$~, by
\begin{equation}
  \left<\vec p^{\,2}\,\right>=\int\,d^3p\,\vec p^{\,2}\,\phi(\vec p\,)=
  {3\over 2} p_F^2. \label{eq:pF}
\end{equation}
where the momentum probability distribution function is given by
\begin{equation}
  \phi(\vec p\,)=\left(2\over \sqrt{\pi}p_F\right)^3
  \,e^{-\vec p^{\,2}/p_F^2}. \label{eq:prob}
\end{equation}
They calculated the lhs of Eq.~(\ref{eq:pF}) to obtain $p_F$ by using the 
Gaussian form of the wave function and then derive the relations between
physical quantities and this $p_F$.\cite{Kim} 
We have the radial wave function given by Eq.~(\ref{eq:waver}) 
different from a Gaussian one and hence should have relations among physical
qunatities and our parameters, $a$, $b$, $\alpha_s$, and $m_c$, independent 
of $p_F$ and hence we may calculate the lhs of Eq.~(\ref{eq:pF}) to check 
if our calculation gives the value similar to other calculations.
Our value of the lhs of Eq.~(\ref{eq:pF}) gives for 
$\left<\vec p^{\,2}\,\right>$ of the $B$ meson at the first order
\begin{equation}
  \left<\vec p^{\,2}\,\right>=0.560~{\rm GeV}^2, \label{eq:pFvalue}
\end{equation}
and the second order gives $\left<\vec p^{\,2}\,\right>=0.562~{\rm GeV}^2$, 
which should be compared with the latest values $p_F=0.5-0.6$ GeV calculated
in \cite{Kim} which correspond to $\left<\vec p^{\,2}\,\right>=0.375-0.540
~{\rm GeV}^2$.

xi)~When one takes an overall look at the calculated masses, the negative 
component contributions, $n_i/M_0$, are relatively large for both scalar
states, $0^\pm$, at the first as well as second order even though they become
very small for higher spin states. Positive components constantly contribute
to all states. When one compares the first order with the second order 
calculations, one can not conclude that the second order is better than
the first as a whole although higher spin states of $D$ and $B$ are largely 
improved at the second order. This conclusion may be also supported by the 
comparison of the first and second order calculations of the parameters,
$\lambda_1$, $\lambda_2$, and $\bar \Lambda$ with other calculations.
\cite{INC3} In order to incorporate the second order effects properly, one may
need to introduce a different potential and/or method from ours as mentioned 
in v) in this section.

We have used the following algorithms to numerically calculate the heavy meson
masses, Gauss-Hermite quadrature to evaluate integrals and the tridiagonal 
QL implicit method to determine the eigenvalues and eigenvectors of a finite 
dimensional real matrix.\cite{Press}

\section{Comments and Discussions}
\label{sec:com}
In this paper, we have calculated heavy meson masses like $D_{(s)}$, 
$D_{(s)}^*$, $B_{(s)}$, and $B_{(s)}^*$ based on the formulation proposed 
before,\cite{Matsuki1} which
develops the perturbation potential theory in terms of inverse
power of a heavy quark mass. The first and second order calculations of masses
are in good agreement with the experimental data except for the higher spin 
states even though the second order calculation does not much improve the 
first order. The first order calculation of the HQET quantities, 
$\lambda_1$, $\lambda_2$, and $\bar \Lambda$, are also in good agreement with 
the other calculations.\cite{INC3} A new study on the HQET introduced the 
Fermi momentum, $p_F$, to obtain other physical quantities.\cite{Kim} 
Although we have not had $p_F$ in our mind at the begining of this study, 
the obtained value given by Eq.~(\ref{eq:pFvalue}) is in good agreement 
with what they have obtained.\cite{Kim}

We have also found a new symmetry already mentioned in the paper 
\cite{Matsuki1} and realized by the operator, Eq.~(\ref{eq:ok}), 
\[
  -\beta_q\left(\vec\Sigma_q\cdot\vec\ell+1\right),
\]
which is always present when one considers a centrally symmetric
potential model for two particles or when one takes a rest frame limit of
a general relativistic form of the wave function and is related to a 
light quark spin structure, i.e., $y_{j\,m}^k(\Omega)$. That is, this is quite
a general symmetry, not a special feature peculiar to our potential model.

One can easily see degeneracy among the lowest lying pseudoscalar
and vector states as follows. Define
\begin{equation}
  \left| P \right\rangle =
  U^{-1}_c\left(\matrix{0&\Psi _{0\;0}^{-1}}\right),
  \quad \left| {V,\;\lambda } \right\rangle =
  U^{-1}_c\left(\matrix{0&\Psi _{1\;\lambda}^{-1}}\right),
\end{equation}
where the inverse of the charge conjugate operator, $U^{-1}_c$, 
is defined in the Appendix \ref{app:schrodinger} Eqs.~(\ref{eq:Hfwt}), 
$\Psi _{j\;m}^k$ is an eigenfunction obtained in the last chapter,
The explicit forms of these wave functions are given in the 
Appendix \ref{app:0thsol}, Eqs.~(\ref{eq:app:jp0}, \ref{eq:app:jp1}).
and the quantum number $k$ can take only $\pm j$, or $\pm \left( {j+1} 
\right)$. Assigning these states to $D$ mesons, one can have
\begin{equation}
  \left| P \right> =\left|D^\pm \right>,\;{\rm or}\;\left|D^0\right>,
  \quad \left| {V,\;\lambda } \right> =\left|D^*\right>.
\end{equation}
Since these states have the same quantum number $k=-1$,
these have the same masses as well as the same wave functions
up to the zero-th order calculation in $1/m_Q$. That is, the degeneracy 
among these states is simply the result of the special property of the 
eigenvalue equation. Higher order corrections can be obtained by developing 
perturbation of energy and wave function for each state in terms of 
$\Lambda_{\rm QCD}/m_Q$ as given by Eqs.~(\ref{eq:mass}, \ref{eq:wf}),
\begin{eqnarray*}
  \tilde E&\equiv& E-m_Q=E^\ell_0+E^\ell_1+E^\ell_2+\ldots,\\
  \psi_{FWT}&=&\psi^\ell_0+\psi^\ell_1+\psi^\ell_2+\ldots,
\end{eqnarray*}

Finally we would like to discuss qualitative features of form factors/
Isgur-Wise functions. Let us think about to calculate form factors
for semileptonic decay of $B$ meson into $D$. Taking a simple form for the 
lowest lying wave function both for $B$ and $D$ as 
\[
  \Psi^{1S} \sim e^{-b^2\;r^2/2},
\]
where a parameter $b$ is determined by a variational principle,
$\delta(\Psi^{1S\;\dagger}H\Psi^{1S})=0$. Then form factors are 
given by
\[
  F(q^2) \sim \exp\left[ {\rm const.}\;{\tilde E}^2\; (q^2-q^2_{\max})
  \right], \;{\rm or}\quad \xi(\omega) \sim \exp\left[ {\rm const.}\;
  {\tilde E}^2\; (\omega-1)\right],
\]
where
\[
  q^2=(p_B-p_D)^2,\quad \omega=v_B \cdot v_D,\quad
  q^2_{\max}=(m_B-m_D)^2 \leftrightarrow \omega_{\max}=1,
\]
with $v_{B, D}^\mu$ being four-velocity of $B$ and/or $D$ meson.
This means behavior of form factors strongly depends on an eigenvalue, 
${\tilde E}=E-m_Q$ of the eigenvalue equation, Eq.~(\ref{0th:2}), which 
is often called "inertia" parameter $\bar{\Lambda}_q$ when 
$m_Q\rightarrow\infty$. This quantity $\tilde E$ does not depend on any 
heavy quark 
properties at the zero-th order. This result also means that the slope at 
the origin of the Isgur-Wise function includes the term proportional to 
$\tilde E^2$. The constant term ($-1/4$ like the Bjorken limit \cite{Bjorken})
for this slope should be given by a kinematical factor multiplied with the 
above expression.

To conclude, although there have been various relativistic bound state 
equations proposed so far, nobody has yet determined what the most preferable
is.  We believe that our approach presented here must be a promising candidate.
\acknowledgments
The authors would like to thank Koichi Seo for critical comments on our
paper. One of the authors (TM) would like to thank the theory group of 
Institute for Nuclear Study for a warm hospitality where a part of this 
work has been done.

\appendix
\section{Schr\"odinger Equation}
\label{app:schrodinger}
In order to derive Eq.\ (\ref{eq:schrod1}) or Eq.\ ({\ref{eq:schrod2}),
we need to calculate the expectation value,
\begin{equation}
  \left<\psi\right|\left({\cal H}_0+{\cal H}_{\rm int}-E\right)
  \left|\psi\right>, \label{eq:exp}
\end{equation}
by using the equal-time anti-commutation relations among quark fields,
\begin{eqnarray}
  \left\{q_\alpha^c (x),\,q_\beta^{c\,\dagger} (x')\right\}
  _{x_0=x_0^\prime}=\delta_{\alpha\,\beta}\delta^3
  \left(\vec x-\vec x^\prime\right), \nonumber \\
  \left\{Q_\alpha (x),\,Q_\beta^\dagger (x')\right\}
  _{x_0=x_0^\prime}=\delta_{\alpha\,\beta}\delta^3
  \left(\vec x-\vec x^\prime\right). \nonumber
\end{eqnarray}
Since the wave function, $\psi_{\alpha\,\beta}(x-y)$ defined in 
Eq.\ (\ref{eq:state}), is normalized to be constant,
what we need to do is to variate Eq.\ (\ref{eq:exp}) in
terms of $\psi_{\alpha\,\beta}(x-y)$
and to set it equal to zero. Then we can easily obtain Eq.\ (\ref{eq:schrod1})
with the effective hamiltonian given by Eq.\ (\ref{eq:hamil}).
In the course of this derivation, it appears to be clear that 
$\vec p$ only operates on wave functions while $\vec q$ only on
potentials.

To derive the FWT and charge conjugate transformed Schr\"odinger equation 
given by Eq.\ (\ref{eq:schrod2}), we have used the following definitions:
\begin{mathletters}
\label{eq:Hfwt}
\begin{eqnarray}
    &&H_{FWT}=U_c\,U_{FWT}\left(p^\prime_Q\right)\,H\,U^{-1}_{FWT}
    \left(p_Q\right)
    \,U_c^{-1}, \quad \psi _{FWT}=U_c\,U_{FWT}\left( p_Q \right)\psi,\\
    &&U_{FWT}\left( p \right)=\exp \left( {W\left( p \right)
    \,\vec \gamma_Q \cdot \vec {\hat p}} \right)=\cos W+\vec \gamma_Q\cdot
    \vec{\hat p}\,\sin W,\\
    &&\vec {\hat p}={{\vec p} \over p},\quad \tan W\left( p \right)
    ={p \over {m_Q+E}},\quad E=\sqrt{{\vec p~}^2+m_Q^2}, \\
    &&U_c=i\,\gamma^0_Q\gamma^2_Q=-U_c^{-1}.
\end{eqnarray}
\end{mathletters}
Note that the argument of the FWT transformation, $U_{FWT}$, operating on 
a hamiltonian from left is different from the right-operating one, since an 
outgoing momentum, ${\vec p_Q~}^\prime$, is different from an incoming
one, $\vec p_Q$. However, here in our study we work in a configuration space 
which means momenta are nothing but the derivative operators and when we write
them differently, for instance as $\vec p_Q$ and ${\vec p_Q\,}^\prime$, 
their expressions are reminders of their momentum representation. 
Hence although 
the arguments of $U_{FWT}$ and $U_{FWT}^{-1}$ look different $\vec p_Q$ 
and ${\vec p_Q\,}^\prime$ are the same derivative operator, 
$-i\vec \nabla$. Here the difference between
$\vec p_Q$ and ${\vec p_Q\,}^\prime$ is $\vec q$ which operates only on
potentials and gives nonvanishing results.
The FWT transformation is introduced so that a heavy
quark inside a heavy meson be treated as a non-relativistic color source. The
charge conjugation operator, $U_c$, is introduced so that gamma matrices of
a light anti-quark is multiplied from left while those of a heavy quark
from right, which is expressed by using a notation $\otimes$.

To derive Eqs.\ (\ref{eq:hameqs}), we need to first expand $H_{\rm FWT}$ 
in $1/m_Q$ and then take into account the following
properties of a charge conjugation operator, $U_c=i~\gamma^0_Q\gamma^2_Q$, 
to obtain the final expressions,
\begin{equation}
  U_c^{-1}\gamma_Q^\mu\, U_c =-{\gamma_Q^\mu}^T,\\
\end{equation}
i.e.,
\begin{eqnarray}
  &&U_c\,\beta_Q\, U_c^{-1} = -{\beta_Q}^T,\quad
  U_c\,\vec\alpha_Q\, U_c^{-1} = -{\vec\alpha_Q}^T,\quad
  U_c\,\vec\Sigma_Q\, U_c^{-1} = -{\vec\Sigma_Q}^T,\nonumber \\
  &&U_c\,\vec\gamma_Q\, U_c^{-1} = -{\vec\gamma_Q}^T,\quad
  U_c\,\gamma_Q^5\, U_c^{-1} = -{\gamma_Q^5}^T,
\end{eqnarray}
where a superscript $T$ means its transposed matrix.
\section{Derivation of Perturbation}
\label{app:pert}
In this Appendix we will follow the paper \cite{Matsuki1} to
derive perturbative Schr\"odinger equation for a hamiltonian given by 
Eq.~(\ref{eq:hamil}) for consistency. Following that paper, we will give 
an equation at each order by using the explicit interaction terms 
given by Eqs.~(\ref{eq:hameqs}). Here we quote the same equations given in the
former sections for clarification of derivation. The fundamental Schr\"odinger
equation is given by
\begin{equation}
  \left(H_{FWT}-m_Q\right)~\otimes\psi _{FWT}
  =\tilde E\;\psi _{FWT},\label{eq:app:schrod0}
\end{equation}
and expansion of each quantity in $1/m_Q$ is given by
\begin{mathletters}
\label{eq:app:schrod}
\begin{eqnarray}
  H_{FWT}-m_Q &=& H_{-1}+H_0+H_1+H_2+\cdots ,\label{eq:app:hamil} \\
  \tilde E&=&E^\ell_0+E^\ell_1+E^\ell_2+\ldots,\label{eq:app:mass} \\
  \psi_{FWT}&=&\psi^\ell_0+\psi^\ell_1+\psi^\ell_2+\ldots,\label{eq:app:wf}
\end{eqnarray}
\end{mathletters}
With a help of projection operators defined by
\begin{equation}
    \Lambda_\pm = {{1\pm \beta_Q}\over 2},
\end{equation}
we will derive the Schr\"odinger equation at each order.
\subsection{-1st order }
\label{app:subsec:-1st}
From Eqs.~(\ref{eq:app:schrod0}) and (\ref{eq:app:schrod}), 
the -1st order Schr\"odinger equation in $1/m_Q$ is given by
\begin{equation}
    -2m_Q\Lambda_+\otimes\psi^\ell_0=0,
\end{equation}
which means
\begin{equation}
    \psi^\ell_0=\Lambda_-\otimes\psi^\ell_0.
\end{equation}
Remember that matrices of a heavy quark should be multiplied from right.
That is, the zero-th order wave function has only a positive component of
the heavy quark sector and is given by
\begin{equation}
  \psi _0^\ell=\Psi_\ell^+=
  \left( {\matrix{0&{f_\ell\left(\vec r\right)}\cr 
  0&{g_\ell\left(\vec r \right)}\cr }} \right),\label{eq:app:psi}
\end{equation}
where $f_\ell$ and $g_\ell$ are 2 by 2 matrices. More explicit 
form of this wave function is given in the Appendix \ref{app:0thsol}.
\subsection{Zero-th order }
\label{app:subsec:0-th}
The zero-th order equation is given by
\begin{equation}
    -2m_Q\Lambda _+\otimes\psi _1^\ell+H_0\otimes\psi _0^\ell
    =E_0^\ell\psi _0^\ell.
\end{equation}
Multiplying projection operators, $\Lambda_\pm$, from right, respectively,
we obtain
\begin{eqnarray}
  &&H_0\Lambda _-\otimes\psi _0^\ell=E_0^\ell\psi 
_0^\ell,\label{eq:app:0th:1}\\
  &&-2m_Q\Lambda _+\otimes\psi _1^\ell+H_0\Lambda _+\otimes\psi _0^\ell
  =0,\label{eq:app:0th:2}
\end{eqnarray}
whose explicit forms are given by Eqs.\ (\ref{0th:1}) and (\ref{0th:2}),
where use has been made of 
\[
  \Lambda_+\otimes\psi_0^\ell=0.
\]
Detailed analysis of Eq.\ (\ref{eq:app:0th:1}) is given in the Appendix 
\ref{app:0thsol}. When one expands
$\Lambda_+$ components of $\psi _1^\ell$ in terms of the eigenfunctions,
\begin{equation}
  \Psi_\ell^-=\left( {\matrix{{f_\ell\left(\vec r\right)}&0\cr 
  {g_\ell\left(\vec r\right)}&0\cr }} \right).\label{eq:app:Psi-}
\end{equation}
like
\begin{equation}
    \Lambda _+\otimes\psi _1^\ell=\sum\limits_{\ell\,'}
    {c_{1-}^{\ell \,\ell\,'} \Psi _{\ell\,'}^-},
\end{equation}
one can solve Eq.\ (\ref{eq:app:0th:2}) to obtain coefficients, 
$c^{\ell\,\ell\,'}_{1-}$, as
\begin{equation}
  c_{1-}^{\ell \,\ell\,'} ={1\over 2m_Q}\left<{\Psi^-_{\ell\,'}}\right|
  \Lambda _-H_0\Lambda_+\left|\Psi^+_\ell\right>,\label{eq:app:c1-}
\end{equation}
whose explicit form is given by Eq.\ (\ref{eq:c1-}).
Here in this paper the eigenfunctions, $\Psi_\ell^\pm$ are normalized to be 1,
\begin{equation}
     \left<{\Psi ^\alpha_\ell}\right.\left|\Psi^\beta_{\ell\,'}\right>
     =\delta_{\ell\,{\ell\,'}}\delta^{\alpha\,\beta}
     \quad {\rm for}\quad\alpha,\,\beta=+\;{\rm or}\,-.\label{eq:app:normal0}
\end{equation}
\subsection{1st order }
\label{app:subsec:1st}
The 1st order equation is given by
\begin{equation}
    -2m_Q\Lambda _+\otimes\psi _2^\ell+H_0\otimes\psi _1^\ell+
    H_1\otimes\psi _0^\ell =E_0^\ell\psi _1^\ell+E_1^\ell\psi _0^\ell.
\end{equation}
As in the above case, multiplying projection operators from right,
we obtain
\begin{eqnarray}
  &&H_0\Lambda _-\otimes\psi _1^\ell+H_1\Lambda _-\otimes\psi _0^\ell
  =E_0^\ell\Lambda _-\otimes\psi _1^\ell+E_1^\ell\psi 
  _0^\ell,\label{eq:app:1st:1}\\
  &&-2m_Q\Lambda _+\otimes\psi _2^\ell+H_0\Lambda _+\otimes\psi _1^\ell+
  H_1\Lambda _+\otimes\psi _0^\ell
  =E_0^\ell\Lambda _+\otimes\psi _1^\ell.\label{eq:app:1st:2}
\end{eqnarray}
The first equation, Eq.\ (\ref{eq:app:1st:1}), can be solved like in the
ordinary perturbation theory of quantum mechanics.
First expanding $\psi_1^k$ in terms of $\Psi_k^\pm$ as
\begin{equation}
  \psi_1^\ell = \sum\limits_\ell\,\left(c_{1+}^{\ell \,\ell\,'}
  \Psi _{\ell\,'} ^+ + c_{1-}^{\ell \,\ell\,'}\Psi _{\ell\,'} ^- \right),
\end{equation}
and next taking the inner product of the whole Eq.\ (\ref{eq:app:1st:1})
with $\left<\Psi _k ^+\right|$, one obtains
\begin{eqnarray}
  &&E_0^k\,c_{1+}^{\ell \,k}+
  \sum\limits_{\ell\,'}\,c_{1-}^{\ell \,{\ell\,'}}\left<
  {\Psi _k ^+}\right| \Lambda _+H_0 \Lambda _-\left| \Psi _{\ell\,'} ^-\right>
  +\left<{\Psi _k ^+}\right|
  \Lambda _-H_1 \Lambda _-\left| \Psi _\ell ^+\right> \nonumber \\
  &&\qquad =
  E_0^\ell\,c_{1+}^{\ell \,k}+E_1^k\,\delta_{{\ell\,k}},\label{eq:app:1st:3}
\end{eqnarray} 
where we have used the orthogonality condition, Eq.\ (\ref{eq:app:normal0}),
and the lowest Schr\"odinger equation, Eq.\ (\ref{eq:app:0th:1}), to obtain
the first term of the lhs and two terms of the rhs of 
Eq.~(\ref{eq:app:1st:3}). 
When one sets $k=\ell$ in Eq.\ (\ref{eq:app:1st:3}), one obtains
\begin{equation}
  E_1^\ell=\sum\limits_{\ell\,'}\,c_{1-}^{\ell \,{\ell\,'}}
  \left<\Psi _\ell ^+ \right|\Lambda _+H_0\Lambda _-\left|
  \Psi _{\ell\,'} ^-\right>+\left<{\Psi _\ell ^+}\right| \Lambda _-H_1 
  \Lambda _-\left|\Psi _\ell ^+\right>,
\end{equation}
which gives the first order perturbation correction to the mass when
one calculates matrix elements of the rhs among eigenfunctions,
$\Psi^\pm_k$, like in the ordinary perturbation of quantum mechanics. 
When one sets $k\ne \ell$ in Eq.\ (\ref{eq:app:1st:3}), one obtains
\begin{eqnarray}
  c_{1+}^{\ell \,k}={1\over E_0^\ell-E_0^k}\left[
  \sum\limits_{\ell\,'}\,c_{1-}^{\ell \,{\ell\,'}}
  \left<\Psi _k ^+ \right|\Lambda _+H_0\Lambda _-\left|
  \Psi _{\ell\,'} ^-\right>+  \left<{\Psi _k ^+}\right| \Lambda _-H_1 
  \Lambda _-\left|\Psi _\ell ^+\right>\right].\label{eq:app:c1+}
\end{eqnarray}
The coefficient $c_{1+}^{k\,k}$ cannot be determined by the above equation,
which can be derived by calculating
a normalization of the total wave function up to the first order,
\begin{equation}
    \left<{\psi^\ell}\right.\left|\psi^{\ell\,'}\right>
    =\delta_{\ell\,{\ell\,'}},\label{eq:app:normal}
\end{equation}
giving
\begin{equation}
  c_{1+}^{k\,k}=0.\label{eq:app:c1+kk}
\end{equation}
This completes the solution for $\psi _1^\ell$ since $\Lambda_-\psi^\ell_1$,
or $c_{1-}^{\ell\,{\ell\,'}}$, is obtained in the last chapter, 
Eq.~(\ref{eq:app:c1-}).
The definition of the normalization, Eq.\ (\ref{eq:app:normal}), is already
mentioned in the main text below Eq.\ (\ref{eq:normpsi} and 
hence we will not repeat that argument here.
This way of solving Eq.\ (\ref{eq:app:1st:1}) is unique and we will 
use this method below to solve similar equations appearing the subsection
\ref{subsec:2nd} as well.

Eq.\ (\ref{eq:app:1st:2}) gives a $\Lambda_+$ component of $\psi^\ell_2$
as in the case of $\Lambda_+\otimes\psi^\ell_1$ in the subsection 
\ref{subsec:0-th}, i.e., setting
\begin{equation}
    \Lambda _+\otimes\psi _2^\ell=\sum\limits_{\ell\,'}
    {c_{2-}^{\ell \,{\ell\,'}}\Psi _{\ell\,'} ^-},
\end{equation}
one obtains coefficients, $c^{\ell\,{\ell\,'}}_{2-}$, as
\begin{equation}
    c_{2-}^{\ell \,{\ell\,'}}={1 \over {2m_Q}}
    \left<{\Psi^- _{\ell\,'}}\right.\left|
    \left(\left(H_0-E^\ell_0\right)\Lambda _+\otimes\psi^\ell_1 
    +H_1\Lambda _+\otimes\psi^\ell_0 \right)\right>.\label{eq:app:c2-}
\end{equation}
\subsection{2nd order }
\label{app:subsec:2nd}
The 2nd order equation is given by
\begin{equation}
    -2m_Q\Lambda _+\otimes\psi _3^\ell+H_0\otimes\psi _2^\ell+
    H_1\otimes\psi _1^\ell+H_2\otimes\psi _0^\ell
    =E_0^\ell\psi _2^\ell+E_1^\ell\psi _1^\ell+E_2^\ell\psi_0^\ell.
\end{equation}
As in the above cases, multiplying projection operators from right we obtain
\begin{eqnarray}
  &H_0\Lambda _-\otimes\psi _2^\ell+H_1\Lambda _-\otimes\psi _1^\ell+
  H_2\Lambda_-\otimes\psi _0^\ell
  =E_0^\ell\Lambda _-\otimes\psi _2^\ell+E_1^\ell\Lambda _-\otimes\psi _1^\ell
  +E_2^\ell\psi_0^\ell,&\label{eq:app:2nd:1}\\
  &-2m_Q\Lambda _+\otimes\psi _3^\ell+H_0\Lambda _+\otimes\psi _2^\ell+
  H_1\Lambda _+\otimes\psi _1^\ell
  +H_2\Lambda _+\otimes\psi _0^\ell & \nonumber \\
  &=E_0^\ell\Lambda _+\otimes\psi _2^\ell
  +E_1^\ell\Lambda _+\otimes\psi _1^\ell.& \label{eq:app:2nd:2}
\end{eqnarray}
Again to solve the first equation, Eq.\ (\ref{eq:app:2nd:1}), 
first expanding $\psi_2^\ell$ in terms of $\Psi_\ell^\pm$ as
\begin{equation}
  \psi_2^\ell = \sum\limits_\ell\,\left(c_{2+}^{\ell \,{\ell\,'}}
  \Psi _{\ell\,'} ^+ +  c_{2-}^{\ell \,{\ell\,'}}\Psi _{\ell\,'} ^-\right),
\end{equation}
and next taking the inner product of the whole Eq.~(\ref{eq:app:2nd:1})
with $\left<\Psi _k ^+\right|$, one obtains
\begin{eqnarray}
  &&E_0^k\,c_{2+}^{\ell \,k}+
  \sum\limits_{\ell\,'}\,c_{2-}^{\ell \,{\ell\,'}}\left<
  \Psi _k ^+ \right|\Lambda _+H_0 \Lambda _-\left|\Psi _{\ell\,'} ^-\right>+
  \left<\Psi _k ^+\right| H_1 
  \Lambda _-\left.\otimes\psi _1 ^\ell\right>
  \nonumber \\
  &&\quad + \left<{\Psi _k ^+}\right| \Lambda _-H_2 \Lambda _-\left|
  \Psi _\ell ^+\right>=E_0^\ell\,c_{2+}^{\ell \,k}+E_1^\ell\,c_{1+}^{\ell \,k}
  +E_2^\ell\,\delta_{\ell\,k},\label{eq:app:2nd:3}
\end{eqnarray} 
where we have used the orthogonality condition, Eq.\ (\ref{eq:app:normal0}),
and the lowest Schr\"odinger equation, Eq.\ (\ref{eq:app:0th:1}), to obtain
the first term of the lhs and three terms of the rhs of 
Eq.\ (\ref{eq:app:2nd:3}). When one sets $k=\ell$ 
in Eq.~(\ref{eq:app:2nd:3}), since $c_{1+}^{\ell\,\ell}=0$ one obtains
\begin{equation}
  E_2^\ell=
  \sum\limits_{\ell\,'}\,c_{2-}^{\ell \,{\ell\,'}}\left<
  {\Psi _\ell ^+}\right| \Lambda _+H_0 \Lambda _-\left|\Psi _{\ell\,'} ^-
  \right>+\left< {\Psi _\ell ^+}\right| H_1\Lambda _- \left.\otimes
  \psi _1 ^\ell\right>+\left<{\Psi _\ell ^+}\right| \Lambda _-H_2 \Lambda _-
  \left|\Psi _\ell ^+\right>,
\end{equation} 
which gives the second order perturbation corrections to the mass when
one calculates matrix elements of the rhs among eigenfunctions.
When one sets $k\ne \ell$ in Eq.\ (\ref{eq:app:2nd:3}), one obtains
\begin{eqnarray}
  c_{2+}^{\ell\,k}&=&{1\over E_0^\ell-E_0^k}\left[
  \sum\limits_{\ell\,'}\,c_{2-}^{\ell \,{\ell\,'}}\left<
  \Psi _k ^+\right| \Lambda _+H_0 \Lambda _-\left|\Psi _{\ell\,'} ^-\right>+
  \left< \Psi _k ^+\right| H_1\Lambda _- \left.\otimes\psi _1 ^\ell\right>
  \right. \nonumber \\
  &&\left.+ 
  \left<\Psi _k ^+\right| \Lambda _-H_2
  \Lambda _-\left|\Psi _\ell ^+\right>
  -E_1^\ell\,c_{1+}^{\ell\,k}\right].
\end{eqnarray}
The coefficient $c_{2+}^{k\,k}$ can be derived by calculating
a normalization of the total wave function up to the second order,
Eq.\ (\ref{eq:app:normal}), which gives
\begin{equation}
  c_{2+}^{k\,k}=
  -{1\over 2}\sum\limits_\ell\left(\left| c_{1+}^{k\,\ell}\right|^2
  +\left| c_{1-}^{k\,\ell}\right|^2 \right).
\end{equation}
This completes the solution for $\psi _2^k$ since $\Lambda_-\psi^k_2$,
or $c_{2-}^{\ell\,{\ell\,'}}$, is obtained in the last chapter, 
Eq.~(\ref{eq:app:c2-}).

Although we do not use, Eq.\ (\ref{eq:app:2nd:2}) gives a $\Lambda_+$ 
component of $\psi^\ell_3$ as in the cases of $\Lambda_+\otimes\psi^\ell_1$ 
and $\Lambda_+\otimes\psi^\ell_2$ given in the subsections \ref{subsec:0-th}
and \ref{subsec:1st}, i.e., setting
\begin{equation}
    \Lambda _+\otimes\psi _3^\ell=\sum\limits_{\ell\,'}
    {c_{3-}^{\ell \,{\ell\,'}}\Psi _{\ell\,'}^-},
\end{equation}
one obtains coefficients, $c^{\ell\,{\ell\,'}}_{3-}$, as
given by Eq.\ (\ref{eq:c3-})
\section{Zero-th Order Solution}
\label{app:0thsol}
There have been a couple of trials to solve 
Eq.\ (\ref{0th:1}).\cite{Critch,Rein,Morii1}
In order to solve the lowest non-trivial eigenvalue equation given by
Eq.\ (\ref{0th:1}), i.e.,
\begin{equation}
  \left[\,\vec\alpha_q\cdot\vec p+\beta_q\left(m_q+S\right)+V\,\right]
  \otimes\psi _0^\ell=E_0^\ell\psi _0^\ell,\label{app:0th}
\end{equation}
we summarize and refresh the previous results. First
we need to introduce the so-called vector spherical harmonics
which are defined by
\begin{eqnarray}
  \vec Y_{j\,m}^{(\rm L)} &=& -\vec n\,Y_j^m, \\
  \vec Y_{j\,m}^{(\rm E)} &=& {r \over {\sqrt {j(j+1)}}}\vec \nabla Y_j^m, \\
  \vec Y_{j\,m}^{(\rm M)} &=& -i\vec n\times \vec Y_{j\,m}^{(\rm E)}
  ={{-ir} \over {\sqrt {j(j+1)}}}\vec n\times \vec \nabla Y_j^m,
\end{eqnarray}
where $Y_j^m$ are spherical polynomials (or surface harmonics).
These vector spherical harmonics satisfy the orthogonality condition:
\[
  \int d\Omega {\vec Y_{j\,m}^{(\rm A)}(\Omega)^\dagger }\cdot \vec
  Y_{j^\prime\,m^\prime}^{(\rm B)}(\Omega) =\delta
  _{j\,j^\prime}\delta _{m\,m^\prime}\delta ^{\rm A\,B},
\]
where $d\Omega = \sin\theta d\theta d\phi$. This is nothing but
a set of eigenfunctions for a spin-1 particle. In this paper we
need their spinor representation and also need to unitary-transform
them to obtain the functions, $y^k_{j\,m}$, as
\begin{equation}
  \left( {\matrix{{y_{j\,m}^{-(j+1)}}\cr {y_{j\,m}^j}\cr}} \right)
  =U\left( {\matrix{{Y_j^m}\cr 
  {\vec \sigma \cdot \vec Y_{j\,m}^{(\rm M)}}\cr}} \right),
  \qquad\left( {\matrix{{y_{j\,m}^{j+1}}\cr {y_{j\,m}^{-j}}\cr}} \right)
  =U\left( {\matrix{{\vec \sigma \cdot \vec Y_{j\,m}^{(\rm L)}}\cr
  {\vec \sigma \cdot \vec Y_{j\,m}^{(\rm E)}}\cr}} \right)
\end{equation}
where
\begin{equation}
  U={1 \over {\sqrt {2j+1}}}\left( {\matrix{{\sqrt {j+1}}&{\sqrt j}\cr
  {-\sqrt j}&{\sqrt {j+1}}\cr}} \right). \label{eq:app:u}
\end{equation}
$\vec Y_{j\,m}^{(\rm A)}$ (A=L, M, E) are eigenfunctions of 
${\vec j}^2$ and $j_z$, having the eigenvalues, $j(j+1)$ and $m$. Parities are
assigned as $(-)^{j+1},~(-)^j,~(-)^{j+1}$ for A=L, M, E, respectively, and
$Y_j^m$ has a parity $(-)^j$. Here
\begin{equation}
  {\vec j} = {\vec \ell} + {\vec s_q} +{\vec s_Q},
\end{equation}
and~$\vec s_q=\vec \sigma_q/2$ and $\vec s_Q=\vec \sigma_Q/2$~
are spin operators of light anti-quark and heavy quark, respectively.
$y_{j\,m}^k$ are $2 \times 2$-matrix
eigenfunctions of three operators, $\vec j^{\,2}$,
$j_z$, and $\vec \sigma \cdot \vec \ell$ with eigenvalues, $j(j+1)$,
$m$, and $-(k+1)$, and satisfy
\begin{equation}
  {1 \over 2}{\rm tr}\left( {\int {d\Omega
  \;{y_{j'\,m'}^{k'}}^\dagger \;y_{j\,m}^k}} \right)
  =\delta^{k\,k'}\delta _{j\,j'}\delta _{m\,m'}.\label{eq:app:normaly}
\end{equation}
Here the quantum number $k$ can take only values,
$\pm j$, or $\pm (j+1)$, and $\neq 0$ since $\vec Y_{0\;0}^{(\rm M)}$
does not exit.
The eigenfunction of
the zero-th order hamiltonian depends on three quantum numbers, 
$j$, $m$, and $k$ as described next.
Functions, $\vec Y_{j\,m}^{(\rm A)}$ and $y_{j\,m}^k$, have the following 
correspondence to those defined in Ref.\ \onlinecite{Morii1} as:
\begin{equation}
  \matrix{
  y_{j\, m}^{-(j+1)}\leftrightarrow y_1^{j\, m},&
  y_{j\, m}^j\leftrightarrow y_2^{j\, m},\cr
  y_{j\, m}^{j+1}\leftrightarrow y_+^{j\, m},&
  y_{j\, m}^{-j}\leftrightarrow y_-^{j\, m},\cr}
\end{equation}
and
\begin{equation}
  \matrix{
  \vec Y_{j\,m}^{(\rm L)}\leftrightarrow \vec Y_{j\,(-)}^m,&
  \vec Y_{j\,m}^{(\rm E)}\leftrightarrow \vec Y_{j\,(+)}^m,\cr
  \vec Y_{j\,m}^{(\rm M)}\leftrightarrow \vec X_{j\,j}^m.& \cr}
\end{equation}
The explicit expressions of the first few $y_{j\,m}^k$ are given by
\begin{eqnarray}
  &&y_{00}^{-1}={1\over \sqrt{4\pi}},
  \quad y_{00}^1=-{1\over \sqrt{4\pi}}\;
  \left(\vec\sigma\cdot\vec n\right),\label{eq:app:y00} \\
  &&y_{1m}^{-1}={i\over \sqrt{4\pi}}\sigma^m,\quad
  y_{1m}^1=-{i\over \sqrt{4\pi}}\left(\vec\sigma\cdot\vec n\right)
  \sigma^m,\label{eq:app:y1m}
\end{eqnarray}
where we have used
\begin{equation}
  Y_0^0={1\over \sqrt{4\pi}},\quad Y_1^0=i\sqrt{3\over 4\pi}\cos\theta,\quad
  Y_1^\pm=\mp i\sqrt{3\over 8\pi}\sin\theta\;e^{\pm i\varphi}.
\end{equation}

In order to solve Eq.\ (\ref{app:0th}), one can in general assume the form of
the solution as
\begin{equation}
  \psi _0^\ell=\Psi_\ell^+=\left(\matrix{0&\Psi _{j\,m}^k
  (\vec r)}\right),\label{eq:app:asmsol}
\end{equation}
where $\ell$ stands for all the quantum numbers, $j$, $m$, and $k$ and
\begin{equation}
  \Psi _{j\,m}^k(\vec r)
  =\left( {\matrix{F_k(\vec r)\cr G_k(\vec r)\cr}} \right)\;y_{j\,m}^k
  (\Omega),
\end{equation}
Since the effective lowest order hamiltonian does not include the heavy 
quark matrices, one can exclude the symbol $\otimes$ from Eq.~(\ref{app:0th}).
The form of a radial wave function is in general given by
\begin{equation}
  \left( {\matrix{F_k(\vec r)\cr G_k(\vec r)\cr}} \right)
  =\left( \matrix{f_{1\,k}(r) - f_{2\,k}(r) 
  \left(\vec \sigma \cdot \vec n \right)\cr
  g_{1\,k}(r) - g_{2\,k}(r) 
  \left(\vec \sigma \cdot \vec n \right)\cr} \right).
\end{equation}
Substituting this into Eq.\ (\ref{app:0th}), multiplying 
${y_{j\,m}^k}^\dagger$ from left and using the orthogonality equation for
$y_{j\,m}^k$, Eq.~(\ref{eq:app:normaly}), the simultaneous equations 
for $f_{i\,k}$ and $g_{i\,k}$ are obtained and after some calculations 
the final form of the wave function is determined to be either
\begin{eqnarray}
  &\Psi _{j\,m}^k=\left( {\matrix{{f_{1\,k}\left( r \right)}\cr
  {-g_{2\,k}\left( r \right)\left( {\vec \sigma \cdot \vec n} \right)}\cr
  }} \right)\,y_{j\,m}^k=\left( {\matrix{{f_{1\,k}\left( r \right)
  \,y_{j\,m}^k}\cr{g_{2\,k}\left( r \right)\,y_{j\,m}^{-k}}\cr
  }} \right),& \label{eq:app:sol1}\\
{\rm or}&&\nonumber \\
  &\Psi _{j\,m}^k=\left( {\matrix{{-f_{2\,k}\left( r \right)\left({\vec\sigma
  \cdot \vec n} \right)}\cr {g_{1\,k}\left( r \right)}\cr
  }} \right)\,y_{j\,m}^k=\left( {\matrix{{f_{2\,k}\left( r \right)
  \,y_{j\,m}^{-k}}\cr
  {g_{1\,k}\left( r \right)\,y_{j\,m}^k}\cr
  }} \right)=\left( {\matrix{{f_{2\,k}\left( r \right)}\cr
  {-g_{1\,k}\left( r \right)\left( {\vec \sigma \cdot \vec n} \right)}\cr
  }} \right)\,y_{j\,m}^{-k},&\label{eq:app:sol2}
\end{eqnarray}
where Pauli matrices $\vec\sigma$ are all for a light quark.
Because of Eq.~(\ref{eq:app:sol2}), we generally define the eigenfunction,
$\Psi_{j\,m}^k$, given by Eq.~(\ref{eq:app:sol1}).
Then the reduced Schr\"odinger equation is given by
\begin{equation}
  \left[ i\left( {\partial _r+{1 \over r}}\right)\rho _1
  +{k \over r}\rho _2
  +\left( {m_q+S\left( r \right)} \right)\rho _3
  +V\left( r \right) 
  \right]\Psi _k(r)=E^k_0\,\Psi _k(r),\label{eq:app:schrod1}
\end{equation}
with 
\begin{equation}
  \Psi _k(r)\equiv\left( {\matrix{{f_{1\,k}\left( r \right)}\cr
  {g_{2\,k}\left( r \right)}\cr}} \right).
\end{equation}
Here defined also are
\begin{eqnarray}
  \vec \alpha _q=\left( {\matrix{0&{\vec \sigma _q}\cr{\vec \sigma _q}&0\cr
  }} \right)=\rho _1\,\vec \sigma,\quad\beta _q=\left( {\matrix{1&0\cr
  0&{-1}\cr}} \right)=\rho _3\,1_{2\times 2}, \\
  \rho _1=\left( {\matrix{0&1\cr 1&0\cr}} \right),
  \quad\rho _2=\left( {\matrix{0&{-i}\cr i&0\cr}} \right),
  \quad\rho _3=\left( {\matrix{1&0\cr 0&{-1}\cr}} \right).
\end{eqnarray}
Finally introducing the unitary matrix,
\begin{equation}
  R=\left( {\matrix{1&0\cr 0&{-i}\cr}} \right),
  \quad R^{-1}=\left( {\matrix{1&0\cr 0&i\cr}} \right),
\end{equation}
one can transform eigenvalue equation as well as eigenfunctions into
\begin{equation}
  \left\{ {r\,R \left[ i\left( {\partial _r+{1 \over r}}\right)\rho _1
  +{k \over r}\rho _2 +\left( {m_q+S\left( r \right)} \right)\rho _3
  +V\left( r \right) \right]\,{1 \over r}\,R^{-1}} \right\}
  \Phi _k(r)=E^k_0\,\Phi _k(r),\label{eq:app:schrod2}
\end{equation}
or
\begin{equation}
\left( {\matrix{{m_q+S+V}&{-\partial _r+{k \over r}}\cr
{\partial _r+{k \over r}}&{-m_q-S+V}\cr
}} \right)\left( {\matrix{{u_k\left( r \right)}\cr
{v_k\left( r \right)}\cr
}} \right)=E^k_0\;\left( {\matrix{{u_k\left( r \right)}\cr
{v_k\left( r \right)}\cr
}} \right), \label{eq:app:schrod3}
\end{equation}
with
\begin{equation}
  \Phi _k(r)\equiv\left( {\matrix{{u_k\left( r \right)}\cr
  {v_k\left( r \right)}\cr}} \right)
  =r\,R\,\Psi _k(r)=\left( {\matrix{{r\,f_{1\;k}\left( r \right)}\cr
  {-ir\,g_{2\;k}\left( r \right)}\cr}} \right).\label{eq:app:phi}
\end{equation}
Then the solution to Eq.\ (\ref{0th:1}) is given by
\begin{equation}
  \Psi _{j\,m}^k={1\over r}\left( {\matrix{{u_k\left( r \right)}\cr
  {-i\,v_k\left( r \right)\left( {\vec \sigma \cdot \vec n} \right)}\cr
  }} \right)\,y_{j\,m}^k(\Omega)={1\over r}\left( {\matrix{
  {u_k\left( r \right)\,y_{j\,m}^k(\Omega)}\cr
  {i\,v_k\left( r \right)\,y_{j\,m}^{-k}(\Omega)}\cr}}\right).
  \label{eq:app:solution}
\end{equation}
Throughout the above derivation, use has been made of formulae
given in the next Appendix.

In order to see the spin-flavor symmetry in our case, the explicit 
form of each lowest order wave function is given as follows in the case of
$J^P=0^-,~1^-$. That these states are degenerate can be easily seen from
the eigenvalue equation where the eigenvalue $E_0^k$ depends only on the
quantum number $k$ and these states have the same value $k=-1$. 
The pseudo-scalar state ($J^P=0^-$) is given by
\begin{equation}
  \left(\matrix{0&\Psi_{0\,0}^{-1}\cr}\right)={1\over \sqrt{4\pi}\,r}
  \left(\matrix{0&u_{-1}\left( r \right)\cr 0&-i\,v_{-1}\left(r\right)\,
  \left(\vec r\cdot\vec\sigma\right)\cr}\right),\label{eq:app:jp0}
\end{equation}
and the vector state ($J^P=1^-$) is given by
\begin{equation}
  \sum_m{\epsilon^m}\left(\matrix{0&\Psi_{1\,m}^{-1}\cr}\right)={i\over 
\sqrt{4\pi}\,r}
  \left(\matrix{0&u_{-1}\left( r \right)\cr 0&-i\,v_{-1}\left(r\right)\,
  \left(\vec r\cdot\vec\sigma\right)\cr}\right)\,\left(\vec\epsilon\cdot\vec
  \sigma\right),\label{eq:app:jp1}
\end{equation}
where use has been made of Eqs.~(\ref{eq:app:y00}, \ref{eq:app:y1m}).
These are transformed into each other via the unitary rotation
\begin{equation}
  \exp\left({\pi\over 2}\vec\epsilon\cdot\vec\sigma\right).
\end{equation}
Here one has to remember that $\vec\epsilon^{~2}=-1$ and also that
we omit the $U_c^{-1}$ operation on the wave function for simplicity.
Similar degeneracy can be seen for a pair of states with the same value 
of $k$.

\section{Matrix Elements}
\label{app:matrix}
In this appendix, we evaluate matrix elements of the rhs of 
Eqs.\ (\ref{eq:1st:energy}, \ref{eq:2nd:energy}) to obtain mass corrections,
$E_1^k,~E_2^k$, those of the rhs of Eqs.~(\ref{eq:c1+kl}, \ref{eq:c2+kl}, 
\ref{eq:c2+kk}) to obtain the corrections of $\Lambda_-$ components of 
wave functions, $c_{1+}^{\ell\,k},~c_{2+}^{\ell\,k}$,
and to evaluate  Eqs.\ (\ref{eq:c1-}, \ref{eq:c2-}) to obtain the 
corrections of $\Lambda_+$ components of wave functions $\psi^\ell_i$ up to 
the second order ($i=1,\,2$), $c_{1-}^{\ell\,k},~c_{2-}^{\ell\,k}$.

Summarizing $\Lambda_--\Lambda_-$ and/or $\Lambda_--\Lambda_+$ matrix 
elements of the hamiltonian, the following equations
are obtained. The rhs of Eq.~(\ref{eq:c1-}) is given here again as
\begin{equation}
  c_{1-}^{\ell \,k}
  ={1\over 2 m_Q}\left<{\Psi^-_k}\right|H_0^{-+}
  \left|\Psi^+_\ell\right>
  ={1\over 4m_Q}\left<{\Psi^-_k}\right|
  \left[\, {\vec \alpha _q\cdot\vec \alpha _Q+\left( 
  {\vec \alpha _q\cdot \vec n} \right)\left( {\vec \alpha_Q\cdot \vec n} 
  \right)}\, \right]V\left|\Psi^+_\ell\right>,\label{eq:app:c1-2}
\end{equation}
which requires to calculate the zero-th order matrix elements.
Here one must notice the following relation,
\begin{equation}
  c_{1-}^{\ell \,k}={c_{1-}^{k \,\ell}}^*
  ={1\over 2 m_Q}\left<{\Psi^+_\ell}\right|H_0^{+-}
  \left|\Psi^-_k\right>^\dagger.\label{eq:app:c1-3}
\end{equation}
The coefficient $c_{1+}^{\ell\,k}$ is given by, for $\ell\ne k$,
\begin{equation}
  c_{1+}^{\ell\,k}={1 \over E_0^\ell-E_0^k}\left[2m_Q \sum_{\ell\,'}
  c_{1-}^{\ell\,\ell\,'} c_{1-}^{\ell\,'\,k}
  +\left<\Psi_k^+\right|H_1^{--}\left|\Psi_\ell^+\right>\right].
\end{equation}
Using Eq.~(\ref{eq:app:c1-3}), the first order energy correction is given by
\begin{equation}
  E_1^\ell=2m_Q \sum_{\ell\,'}\left|c_{1-}^{\ell\,\ell\,'}\right|^2
  +\left<\Psi_\ell^+\right|H_1^{--}\left|\Psi_\ell^+\right>.\label{eq:app:E1}
\end{equation}
Simplifying Eq.~(\ref{eq:c2-}), one obtains
\begin{eqnarray}
  c_{2-}^{\ell \,k}
  &=&\sum_{\ell^{\prime}}c_{1+}^{\ell\,\ell^{\prime}}
  c_{1-}^{\ell^{\prime}\,k}
  +{1\over 2 m_Q}\left[\left(E_0^k-E_0^\ell \right)
  c_{1-}^{\ell\,k}-2\sum_{\ell^{\prime}}
  c_{1-}^{\ell\,{\ell^{\prime}}}\left<{\Psi^-_k}\right|
  \beta_q\, S\left|\Psi^-_{\ell^{\prime}}\right>
  +\left<{\Psi^-_k}\right|H_1^{-+}\left|\Psi^+_\ell\right>
  \right],\\
  c_{2+}^{\ell \,k}&=&{1\over E_0^\ell-E_0^k}\left[2m_Q\sum_{\ell\,'}
  c_{2-}^{\ell \,\ell\,'}c_{1-}^{\ell\,' \,k}+\sum_{\ell\,'}
  \left(c_{1+}^{\ell \,\ell\,'}\left<\Psi_k^+\right|H_1^{--}\left|
  \Psi_{\ell\,'}^+\right>+c_{1-}^{\ell \,\ell\,'}
  \left<\Psi_k^+\right|H_1^{+-}\left|\Psi_{\ell\,'}^-\right>\right)
  \right. \nonumber \\
  &&\qquad\qquad\qquad \left. +\left<\Psi_k^+\right|H_2^{--}
  \left|\Psi_\ell^+\right>
  -E_1^\ell c_{1+}^{\ell \,k}\right]\qquad {\rm for}\quad\ell\ne k,\\
  E_2^\ell&=&2m_Q\sum_{\ell\,'}
  c_{2-}^{\ell \,\ell\,'}c_{1-}^{\ell\,' \,\ell}+\sum_{\ell\,'}
  \left(c_{1+}^{\ell \,\ell\,'}\left<\Psi_\ell^+\right|H_1^{--}\left|
  \Psi_{\ell\,'}^+\right>+c_{1-}^{\ell \,\ell\,'}
  \left<\Psi_\ell^+\right|H_1^{+-}\left|\Psi_{\ell\,'}^-\right>\right)
  +\left<\Psi_\ell^+\right|H_2^{--}\left|\Psi_\ell^+\right>.\label{eq:app:E2}
\end{eqnarray}
Although it is apparent that $E_1^k$ is always real from 
Eq.~(\ref{eq:app:E1}), it is not clear whether Eq.~(\ref{eq:app:E2}) 
is real or not. We will rewrite Eq.~(\ref{eq:app:E2}) so that reality
of $E_2^k$ is manifest as follows.
\begin{eqnarray}
  E^\ell _2&&=\sum\limits_{\ell\,'} {\sum\limits_{\ell^{\prime\prime}} 
  {c_{1-}^{\ell\,'\;\ell }\;}\left\langle {\Psi _{\ell\,'}^-} \right|
  H_0^{++}\left| {\Psi _{\ell ^{\prime\prime}}^-} \right\rangle }\;
  c_{1-}^{\ell \;\ell ^{\prime\prime}}-E_0^\ell \sum\limits_{\ell\,'} 
  {\left| {c_{1-}^{\ell \;\ell\,'}} \right|}^2+2{\rm Re}\;\sum\limits_{\ell\,'}
  {c_{1-}^{\ell\,'\;\ell }\;}\left\langle {\Psi _{\ell\,'}^-} \right|
  H_1^{-+}\left| {\Psi _{\ell}^+} \right\rangle \nonumber \\
  &&\qquad\qquad\qquad+\sum\limits_{\ell\,'} {{1 \over {E_0^\ell -E_0^{\ell\,'}
  }}}\left[ {4m_Q^2\left| {\sum\limits_{\ell^{\prime\prime}} 
  {c_{1-}^{\ell \;\ell ^{\prime\prime}}c_{1-}^{\ell ^{\prime\prime}\;\ell\,'}}}
  \right|}^2 +\left| {\left\langle {\Psi _\ell ^+} \right|H_1^{--}\left| 
  {\Psi _{\ell\,'}^+} \right\rangle } \right|^2 \right. \nonumber \\
  &&\qquad\qquad\qquad\left. +4m_Q{\rm Re}
  \sum\limits_{\ell ^{\prime\prime}} {c_{1-}^{\ell\;
  \ell^{\prime\prime}}c_{1-}^{\ell ^{\prime\prime}\;\ell\,'}}
  {\left\langle {\Psi _\ell ^+} \right|H_1^{--}\left| {\Psi _{\ell\,'}^+} 
  \right\rangle } \right] +\left\langle {\Psi _\ell ^+} \right|H_2^{--}
  \left| {\Psi _\ell ^+} \right\rangle,\label{eq:app:E22}
\end{eqnarray}
whose expression is apparently real.

In the above derivation, we have used the projected hamiltonian at each order,
$H_i^{\alpha\,\beta}$, which is defined by
\begin{equation}
  \Lambda_\alpha H_i \Lambda_\beta
  \equiv\Lambda_\alpha H_i^{\alpha\,\beta}  \Lambda_\beta
  =\Lambda_\alpha H_i^{\alpha\,\beta}
  =H_i^{\alpha\,\beta}  \Lambda_\beta,
\end{equation}
where $H_i^{\alpha\,\beta}$ is composed of Dirac matrices, $\vec\alpha$,
$\beta$, $\vec\Sigma$, and $\gamma^5$.\cite{Morii3}
\begin{mathletters}
\label{eq:app:ham+-}
\begin{eqnarray}
  H_0^{--}&=&\vec\alpha_q\cdot\vec p+\beta_q\left(m_q+S\right)+V,\\
  H_0^{-+}&=&{1\over 2}\left[\, {\vec \alpha _q\cdot\vec \alpha _Q+\left( 
  {\vec \alpha _q\cdot \vec n} \right)\left( {\vec \alpha_Q\cdot \vec n} 
  \right)}\, \right]V,\\
  H_0^{+-}&=&H_0^{-+},\\
  H_0^{++}&=&\vec\alpha_q\cdot\vec p+\beta_q\left(m_q-S\right)+V,\\
  H_1^{--}&=&{1 \over {2m_Q}}\left\{{\vec p~}^2+V
  \left[ {\left( {\vec \alpha _q\cdot \vec p} \right)-i\,
  \left( {\vec \alpha _q\cdot \vec n} \right)\partial _r} 
  \right]-V^\prime\left[ {i\,\left( {\vec \alpha _q\cdot \vec n} 
  \right)+{1 \over 2}\left( {\vec \alpha _q\cdot \vec \Sigma _Q\times \vec n} 
  \right)} \right]\right. \nonumber \\
  &&\left. -{1\over r}V\left[i\,\left( {\vec \alpha _q\cdot \vec n}
  \right)-{1 \over 2}\left( {\vec \alpha _q\cdot \vec \Sigma _Q\times \vec n} 
  \right) \right] \right\}, \label{eq:app:ham+-:h1--}\\
  H_1^{-+}&=&{1 \over m_Q}\left[-S\beta_q\,\left(\vec\alpha_Q\cdot\vec p\right)
  +{i\over 2}\left(\beta_q\,S^\prime+V^\prime\right)
  \left(\vec\alpha_Q\cdot\vec n\right)\right],\\
  H_1^{+-}&=&{1 \over m_Q}\left[-S\beta_q\,\left(\vec\alpha_Q\cdot\vec p\right)
  +{i\over 2}\left(\beta_q\,S^\prime-V^\prime\right)
  \left(\vec\alpha_Q\cdot\vec n\right)\right],\\
  H_1^{++}&=&-H_1^{--},\\
  H_2^{--}&=&{1 \over {2m_Q^2}}\left\{-\beta_q\,
  \left(\vec p+{1\over 2}\,\vec q\right)^2\,
  S+{1\over 4}\,\Delta V +{1\over 2r}\,\left(\beta_q\,S'-V'\right)\,
  \left(\vec\Sigma_Q\cdot\vec\ell\right) \right\},\\
  H_2^{-+}&=&-{1\over 8m_Q^2}\biggl\{2V\left[ {\left( {\vec \alpha _q\cdot 
  \vec p} \right)-i\,\left( {\vec \alpha _q\cdot \vec n} \right)\partial _r}
  \right]\left(\vec\alpha_Q\cdot\vec p\right)
  -2iV^\prime\left(\vec\alpha_q\cdot\vec n\right)
  \left(\vec\alpha_Q\cdot\vec p\right) \nonumber\\
  &&\left.-iV^\prime\left(\vec\alpha_Q\cdot\vec n\right)
  \left[ {\left({\vec \alpha _q\cdot \vec p}\right)
  -i\,\left( {\vec \alpha _q\cdot \vec n} \right)\partial _r}\right]
  +{1\over r}V^\prime\left({\vec\alpha_q\cdot \vec\ell}\right)
  \gamma_Q^5\right\}+{1\over 8m_Q^2 r}V\,\biggl\{3i\,\left({\vec \alpha _q\cdot
  \vec n}\right) \left(\vec\alpha_Q\cdot\vec p\right)\nonumber \\
  &&\left.+\left[\left(\vec \alpha _q\cdot \vec
  \alpha_Q\right)-2\left({\vec \alpha _q\cdot \vec n}\right)
  \left(\vec\alpha_Q\cdot\vec n\right) \right]\partial_r
  +{1\over r}\left(\vec \alpha _q\cdot \vec \ell\right)\,\gamma_Q^5\right\},\\
  H_2^{+-}&=&H_2^{-+},\\
  H_2^{++}&=&{1 \over {2m_Q^2}}\left\{\beta_q\,
  \left(\vec p+{1\over 2}\,\vec q\right)^2\,
  S+{1\over 4}\,\Delta V -{1\over 2r}\,\left(\beta_q\,S'+V'\right)\,
  \left(\vec\Sigma_Q\cdot\vec\ell\right) \right\}.
\end{eqnarray}
\end{mathletters}
Use has been made of the following formulae for the gamma matrices,
\begin{eqnarray}
  &&\beta_Q\Lambda_\pm=\pm\Lambda_\pm,
  \quad\vec\alpha_Q\Lambda_\pm=\Lambda_\mp\vec\alpha_Q,
  \quad\vec\Sigma_Q\Lambda_\pm=\Lambda_\pm\vec\Sigma_Q,
  \quad\vec\gamma_Q\Lambda_\pm=\Lambda_\mp\vec\gamma_Q
  =\mp\alpha_Q\Lambda_\pm,\nonumber \\
  &&\beta_Q\vec\Sigma_Q\Lambda_\pm=\pm\Lambda_\pm\vec\Sigma_Q,
  \quad\gamma_Q^5 \Lambda_\pm=\Lambda_\mp\gamma_Q^5.
\end{eqnarray}
Matrix elements of interaction terms among eigenfunctions
are calculated below. Degeneracy can be resolved
by heavy quark spin-dependent terms which includes $\vec\alpha_Q$ and/or 
$\vec\Sigma_Q$ dependent terms, i.e., the last terms of $H_1^{--}$ and 
$H_2^{--}$ together with contributions from negative components of the 
wave functions coming from $H_0^{-+}$ and $H_1^{-+}$. 

The formulae necessary for calculating the matrix elements are given below
when operators, 
$\left(\vec \sigma_q \cdot \vec n~\right)$,
$\left(\vec \sigma_q \cdot \vec \ell~\right)$,
$\left(\vec \sigma_q \cdot \vec p~\right)$,
$\left(\vec \sigma_Q \cdot \vec \ell~\right)$, 
$\vec \sigma_q \cdot (\vec \sigma_Q \times \vec n)$, 
$\left(\vec \sigma_q \cdot \vec \sigma_Q~\right)$, 
$\left(\vec \sigma_Q \cdot \vec n~\right)$, and
$\left(\vec \sigma_Q \cdot \vec p~\right)$, 
operate on the function, $y_{j\,m}^k(\Omega)$ or $f(r)\,y_{j\,m}^k(\Omega)$.
The symbol, $\otimes$, is used in the same meaning for $4\times 4$
gamma matrices, i.e., Pauli matrices for a light anti-quark are multiplied 
from left while those for a heavy quark from right.
\begin{eqnarray}
  \left(\vec\sigma_q\cdot\vec n\right)\otimes\,y_{j\,m}^k
  &=&-y_{j\,m}^{-k}, \\
  \left(\vec\sigma_q\cdot\vec\ell\right)\otimes \,y_{j\,m}^k
  &=&-(k+1)\,y_{j\,m}^k, \\
  ~ \nonumber \\
  \left( {\vec \sigma_q \cdot \vec p} \right)\otimes
  \,f\left( r \right)\,y_{j\,m}^k&=&i\left( {\partial _r+{{k+1}
  \over r}} \right)\,f\left( r \right)\,y_{j\,m}^{-k} \nonumber \\
  &=&-i\left( {\partial _r+{{k+1} \over r}} \right)
  \,f\left( r \right)\,\left( {\vec \sigma_q \cdot \vec n}
  \right)\otimes\,y_{j\,m}^k, \\
  ~ \nonumber \\
  \left(\vec\sigma_Q\cdot\vec\ell\right)\otimes
  \left( {\matrix{{y_{j\,m}^{-(j+1)}}\cr
  {y_{j\,m}^j}\cr }} \right)&=&{1 \over {2j+1}}
  \left( {\matrix{{j(2j+3)}&{2\sqrt {j(j+1)}}\cr
  {2\sqrt {j(j+1)}}&{-(2j-1)(j+1)}\cr
  }} \right)\,\left( {\matrix{{y_{j\,m}^{-(j+1)}}\cr
  {y_{j\,m}^j}\cr}} \right), \nonumber \\
  ~ \\
  \left(\vec\sigma_Q\cdot\vec\ell\right)\otimes
  \left( {\matrix{{y_{j\,m}^{j+1}}\cr
  {y_{j\,m}^{-j}}\cr}} \right)&=&\left( {\matrix{{j+2}&0\cr
  0&{-(j-1)}\cr}} \right)\,\left( {\matrix{{y_{j\,m}^{j+1}}\cr
  {y_{j\,m}^{-j}}\cr}} \right), \nonumber \\
  ~ \nonumber \\
  \left( \vec \sigma _q\cdot {\vec \sigma _Q\times \vec n} \right)\otimes
  \left( {\matrix{{y_{j\,m}^{-(j+1)}}\cr {y_{j\,m}^j}\cr }} \right)&=&
  {{2i} \over {2j+1}}\left( {\matrix{{-(j+1)}&{\sqrt {j(j+1)}}\cr
  {\sqrt {j(j+1)}}&{-j}\cr}} \right)\,\left( {\matrix{{y_{j\,m}^{j+1}}\cr
  {y_{j\,m}^{-j}}\cr }} \right), \nonumber \\
  ~ \\
  \left( \vec \sigma _q\cdot {\vec \sigma _Q\times \vec n}\right)\otimes
  \left( {\matrix{{y_{j\,m}^{j+1}}\cr {y_{j\,m}^{-j}}\cr}} \right)&=&
  {{-2i} \over {2j+1}}\left( {\matrix{{-(j+1)}&{\sqrt {j(j+1)}}\cr
  {\sqrt {j(j+1)}}&{-j}\cr }} \right)\,\left( {\matrix{{y_{j\,m}^{-(j+1)}}\cr
  {y_{j\,m}^j}\cr }} \right). \nonumber \\
  ~ \nonumber \\
  \left(\vec\sigma_q\cdot\vec\sigma_Q\right)\otimes
  \left( {\matrix{{y_{j\,m}^{-(j+1)}}\cr 
  {y_{j\,m}^j}\cr }} \right)&=&{1 \over {2j+1}}
  \left( {\matrix{{2j+3}&{-4\sqrt {j(j+1)}}\cr
  {-4\sqrt {j(j+1)}}&{2j-1}\cr
  }} \right)\,\left( {\matrix{{y_{j\,m}^{-(j+1)}}\cr
  {y_{j\,m}^j}\cr}} \right), \nonumber \\
  ~ \\
  \left(\vec\sigma_q\cdot\vec\sigma_Q\right)\otimes
  \left( {\matrix{{y_{j\,m}^{j+1}}\cr
  {y_{j\,m}^{-j}}\cr}} \right)&=&-\left( {\matrix{{y_{j\,m}^{j+1}}\cr
  {y_{j\,m}^{-j}}\cr}} \right), \nonumber \\
  ~ \nonumber \\
  \left(\vec\sigma_Q\cdot\vec n\right)\otimes
  \left( {\matrix{{y_{j\,m}^{-(j+1)}}\cr 
  {y_{j\,m}^j}\cr }} \right)&=&{-1 \over {2j+1}}
  \left( {\matrix{{1}&{-2\sqrt {j(j+1)}}\cr
  {-2\sqrt {j(j+1)}}&{-1}\cr
  }} \right)\,\left( {\matrix{{y_{j\,m}^{-(j+1)}}\cr
  {y_{j\,m}^j}\cr}} \right), \nonumber \\
  ~ \\
  \left(\vec\sigma_Q\cdot\vec n\right)\otimes
  \left( {\matrix{{y_{j\,m}^{j+1}}\cr
  {y_{j\,m}^{-j}}\cr}} \right)&=&{-1 \over {2j+1}}
  \left( {\matrix{{1}&{-2\sqrt {j(j+1)}}\cr
  {-2\sqrt {j(j+1)}}&{-1}\cr
  }} \right)\,\left( {\matrix{{y_{j\,m}^{j+1}}\cr
  {y_{j\,m}^{-j}}\cr}} \right), \nonumber
\end{eqnarray}
\begin{eqnarray}
  &&\left(\vec\sigma_Q\cdot\vec p\right)\otimes
  \left( {\matrix{{f(r)\,y_{j\,m}^{-(j+1)}}\cr 
  {f(r)\,y_{j\,m}^j}\cr }} \right) \nonumber \\
  &&\quad ={i \over {2j+1}}\left( 
  {\matrix{{\partial_r-{j\over r}}&{-2\sqrt {j(j+1)}
  \left(\partial_r+{j+1\over r}\right)}\cr
  {-2\sqrt {j(j+1)}\left(\partial_r-{j\over r}\right)}&
  {-\left(\partial_r+{j+1\over r}\right)}\cr
  }} \right)\,\left( {\matrix{{f(r)\,y_{j\,m}^{-(j+1)}}\cr
  {f(r)\,y_{j\,m}^j}\cr}} \right), \nonumber \\
  ~ \\
  &&\left(\vec\sigma_Q\cdot\vec p\right)\otimes
  \left( {\matrix{{f(r)\,y_{j\,m}^{j+1}}\cr
  {f(r)\,y_{j\,m}^{-j}}\cr}} \right) \nonumber \\
  &&\quad ={i \over {2j+1}}
  \left( {\matrix{{\partial_r+{j+2\over r}}&{-2\sqrt {j(j+1)}
  \left(\partial_r+{j+2\over r}\right)}\cr
  {-2\sqrt {j(j+1)}\left(\partial_r-{j-1\over r}\right)}&
  {-\left(\partial_r-{j-1\over r}\right)}\cr
  }} \right)\,\left( {\matrix{{f(r)\,y_{j\,m}^{j+1}}\cr
  {f(r)\,y_{j\,m}^{-j}}\cr}} \right), \nonumber
\end{eqnarray}
\subsection{$\Lambda_--\Lambda_-$ matrix elements}
\label{app:subsec:++}
\subsubsection{First order terms}
\label{app:subsubsec:1st--}
To calculate Eqs.~(\ref{eq:1st:energy}, \ref{eq:c1+kl}),
one needs the following $\Lambda_--\Lambda_-$ matrix elements,
\begin{eqnarray}
  \left<\Psi^+_{\ell\,'}\right| H^{--}_1 \left|\Psi^+_\ell\right>
  =&&\left<\Psi^+_{\ell\,'}\right|
  {1 \over {2m_Q}}\left\{{\vec p~}^2+V
  \left[ {\left( {\vec \alpha _q\cdot \vec p} \right)-i\,
  \left( {\vec \alpha _q\cdot \vec n} \right)\partial _r} 
  \right]-V^\prime\left( {\vec \alpha _q\cdot \vec \Sigma _Q\times \vec n} 
  \right)\right\}\left|\Psi^+_\ell\right> \nonumber \\
  ={1\over 2}{\rm tr}\int d^3r{\Psi _{j\,m}^{k^\prime}}^\dagger&&\,
  {1 \over {2m_Q}}\left\{ {\vec p~}^2+V
  \left[ {\left( {\vec \alpha _q\cdot \vec p} \right)-i\,
  \left( {\vec \alpha _q\cdot \vec n} \right)\partial _r}\right]
  -V^\prime \left( {\vec \alpha _q\cdot \vec \sigma _Q\times \vec n} 
  \right) \right\}\,\otimes\Psi _{j\,m}^k, \label{eq:app:1st}
\end{eqnarray}
where the sets of quantum numbers are given by $\ell=(j,\,m,\,k)$ and
$\ell\,'=(j,\,m,\,k^\prime)$.
Some simplification occurs because $V(r)\sim 1/r$ hence
$V'=-V/r$. 
Each matrix element of the first order interaction terms is given below.
\begin{eqnarray}
  &&{1\over 2}{\rm tr}\int d^3r\;{\Psi_{j\,m}^k}^\dagger\vec p^{~2}\,
  \otimes\Psi_{j\,m}^k={1\over 2}{\rm tr}\int d^3r\;{\Psi_{j\,m}^k}^\dagger
  \left( {\vec \Sigma _q\cdot \vec p}\right)^2\,\otimes
  \Psi_{j\,m}^k \nonumber \\
  && \nonumber \\
  &&\qquad =\int dr\;\Phi_k^\dagger 
  \left( {\matrix{{-\partial _r^2+{{k\left( {k+1} \right)} \over
  {r^2}}}&0\cr 0&{-\partial _r^2+{{k\left( {k-1} \right)} \over {r^2}}}\cr
  }} \right)\Phi_k,\\
  &&{1\over 2}{\rm tr}\int d^3r\;{\Psi _{j\,m}^k}^\dagger
  V \left[ \left( {\vec \alpha _q\cdot \vec p} \right)-i\,
  \left( {\vec \alpha _q\cdot \vec n} \right)\partial _r
  \right]\,\otimes\Psi _{j\,m}^k \nonumber \\
  &&~ \nonumber \\
  &&\qquad =\int dr\;\Phi _k^\dagger\left(
  {\matrix{0&{-V\left( {2\partial _r-{{k+1} \over r}} \right)}\cr 
  {V\left({2\partial _r+{{k-1} \over r}} \right)}&0\cr }} \right)\Phi_k,
\end{eqnarray}
where $\Phi_k(r)$ is defined by Eq.~(\ref{eq:app:phi}) in the Appendix
\ref{app:0thsol}. Some non-vanishing matrix elements of the last term of
Eq.~(\ref{eq:app:1st}) are given by
\begin{eqnarray}
  &&\left<\Psi^+_{\ell\,'}\right|
  V^\prime \left( {\vec \alpha _q\cdot \vec \Sigma _Q\times \vec n} 
  \right)\,\left|\Psi^+_\ell\right> \nonumber \\
  \nonumber \\
  &&\qquad =\left\{ {\matrix{
  {-2k \over {2k+1}}\int dr\;{\Phi _k^\dagger V'
  \left( {\matrix{0&1\cr 1&0\cr}} \right)
  \Phi _k\quad {\rm for}\;k=-\left( {j+1} \right),\,j,}\cr
  ~\cr
  {2k \over {2k-1}}\int dr\;{\Phi _k^\dagger V'
  \left( {\matrix{0&1\cr 1&0\cr}} \right)
  \Phi _k\quad {\rm for}\;k=j+1,\,-j,}~~~\cr
  ~\cr
  {2{\sqrt {j\left( {j+1} \right)}} \over {2j+1}}\int dr\;\Phi _j^\dagger V'
  \,\left( {\matrix{0&1\cr 1&0\cr }} \right)\Phi _{-\left( 
  {j+1}\right)},\qquad\qquad\cr
  ~\cr
  -{2{\sqrt {j\left( {j+1} \right)}} \over {2j+1}}\int dr\;\Phi _{-j}^\dagger
  V'\,\left({\matrix{0&1\cr 1&0\cr }} \right)\Phi _{j+1},\qquad\qquad\cr
  }} \right.
\end{eqnarray}
and their complex conjugates.
\subsubsection{Second order terms}
\label{app:subsubsec:2nd}
To calculate Eqs.~(\ref{eq:2nd:energy}, \ref{eq:c2+kl}),
one needs the following $\Lambda_--\Lambda_-$ matrix elements,
\begin{eqnarray}
  &&\left<\Psi^+_{\ell\,'}\right| H^{--}_2 \left|\Psi^+_\ell\right>
  =\left<\Psi^+_{\ell\,'}\right|{1 \over {2m_Q^2}}\left\{-\beta_q\,
  \left(\vec p+{1\over 2}\,\vec q\right)^2\,
  S+{1\over 4}\,\Delta V +{1\over 2r}\,\left(\beta_q\,S'-V'\right)\,
  \left(\vec\alpha_Q\cdot\vec\ell~\right) \right\}\left|\Psi^+_\ell\right>,
  \nonumber \\
  && ={1\over 2}{\rm tr}\int d^3r{\Psi _{j\,m}^{k^\prime}}^\dagger\,
  {1 \over {2m_Q^2}}\left\{-\beta_q\,\left(\vec p+{1\over 2}\,\vec q\right)^2
  \,S+{1\over 4}\,\Delta V \right. 
  \left.+{1\over 2r}\,\left(\beta_q\,S'-V'\right)\,
  \left(\vec\sigma_Q\cdot\vec\ell~\right) \right\}\,\otimes\Psi _{j\,m}^k. 
  \label{eq:app:2nd}
\end{eqnarray}
Each matrix element of the second order interaction terms is given below.
\begin{equation}
  {1\over 2}{\rm tr}\int d^3r\;{\Psi_{j\,m}^k}^\dagger\beta _q
  \left({\vec p+{1\over 2}\vec q}\right)^2S
  \,\otimes\Psi_{j\,m}^k 
  =\int dr\;\Phi_k^\dagger 
  \left( {\matrix{{S_+}&0\cr 0&{-S_-}\cr}} \right)\Phi_k,
\end{equation}
where
\begin{equation}
  S_\pm = S\left[ -\partial_r^2+
  {{k\left( {k\pm 1} \right)} \over {r^2}} \right]-
  S' \left(\partial_r-{1 \over {2r}}\right).
\end{equation}
Since $V=-{4\alpha_s /3r}$ and $\Delta {1\over r}
=-4\pi\delta^3(\vec r)$, we need to calculate
\begin{equation}
  {1\over 2}{\rm tr}\int d^3r\;{\Psi_{j\,m}^{k'}}^\dagger\Delta {1\over r}
  \,\otimes\Psi_{j\,m}^k 
  =-\left|\Phi_k(0)\right|^2\,\delta_{k,\,k'}.
\end{equation}
Nonvanishing matrix elements of the last term of Eq.~(\ref{eq:app:2nd}) are
given by
\begin{eqnarray}
  &&\left<\Psi^+_{\ell\,'}\right| 
  {1\over r}\,\left(\beta_q\,S'-V'\right)\,
  \left(\vec\alpha_Q\cdot\vec\ell~\right)\,\left|\Psi^+_\ell\right>
  \nonumber \\
  &&\nonumber ~\\
  &&=\left\{ {\matrix{\int dr\;{\Phi _k^\dagger\left(
  {\matrix{{-{{\left( {k+1} \right)\left( {2k-1} \right)} \over {2k+1}}
  \left({{{S'} \over r}-{{V'} \over r}} \right)}&0\cr 0&{\left( {k-1} 
  \right)\left({{{S'} \over r}+{{V'} \over r}} \right)}\cr }} 
  \right)\Phi _k~~{\rm for}\;k=-\left( {j+1} \right),\,j,}\cr 
  ~\cr
  \int dr\;{\Phi _k^\dagger \left( 
  {\matrix{{\left( {k+1}\right)\left( {{{S'} \over r}-{{V'} \over r}}
  \right)}&0\cr 0&{-{{\left( {k-1}\right)\left( {2k+1} \right)} 
  \over {2k-1}}\left( {{{S'} \over r}+{{V'} \over r}} \right)}\cr }}
  \right)\Phi _k~~{\rm for}\;k=j+1,\,-j,}~~~\cr 
  ~\cr
  {2{\sqrt {j\left( {j+1} \right)}}\over {2j+1}} 
  \int dr\;\Phi _j^\dagger
  \left( {{{S'} \over r}-{{V'} \over r}} \right)\,\left(
  {\matrix{1&0\cr 0&0\cr }} \right)\Phi _{-\left( {j+1} \right)},
  \qquad\qquad\qquad\qquad\cr 
  ~\cr
  -{2{\sqrt {j\left( {j+1} \right)}}\over {2j+1}} \int dr\; 
  \Phi_{-j}^\dagger \left(
  {{{S'} \over r}+{{V'} \over r}} \right)\,\left( {\matrix{0&0\cr 0&1\cr }}
  \right)\Phi _{j+1},\qquad\qquad\qquad\qquad\cr}}\right.
\end{eqnarray}
and their complex conjugates.
\subsection{$\Lambda_--\Lambda_+$ matrix elements}
\label{app:subsec:+-}
\subsubsection{Zero-th order terms }
\label{app:subsubsec:0th-+}
Among the many $\Lambda_--\Lambda_+$ components, that of the rhs of 
Eq.~(\ref{eq:c1-}) is the only matrix element to be needed in the later
calculations at the zero-th order, which is again given here,
\begin{eqnarray}
  \left<{\Psi^-_{\ell\,'}}\right|H_0^{-+}
  \left|\Psi^+_\ell\right>
  &=&\left<{\Psi^-_{\ell\,'}}\right|{1\over 2}
  \left[\, {\vec \alpha _q\cdot\vec \alpha _Q+\left( 
  {\vec \alpha _q\cdot \vec n} \right)\left( {\vec \alpha_Q\cdot \vec n} 
  \right)}\, \right]V\left|\Psi^+_\ell\right>, \nonumber \\
  ~\nonumber \\
  &=&{1\over 2}{\rm tr}\int d^3r{\Psi _{j\,m}^{k^\prime}}^\dagger\,
  {1\over 2}\left[\, {\vec \alpha _q\cdot\vec \sigma _Q+\left( 
  {\vec \alpha _q\cdot \vec n} \right)\left( {\vec \sigma_Q\cdot \vec n} 
  \right)}\, \right]V \otimes\Psi _{j\,m}^k.
\end{eqnarray}
Non-vanishing matrix elements are given by
\begin{eqnarray}
  \left<\Psi_{\ell\,'}^-\right|&&\left(\vec\alpha_q\cdot\vec\alpha_Q
  \right)\,V
  \left|\Psi_\ell^+\right> \nonumber \\
  ~\nonumber \\
  &&=\left\{\matrix{{\int dr\,\Phi_{-k}^\dagger \left(\matrix{
  0&-1\cr {2k-1 \over 2k+1}&0\cr }\right)\,V\Phi_k \quad {\rm for}~
  k=-(j+1),\;j },\cr 
  ~\cr
  {-{4\sqrt{j(j+1)} \over 2j+1}\int dr\,\Phi_{-j}^\dagger \left(\matrix{
  0&0\cr 0&1\cr }\right)\,V\Phi_{-(j+1)}},\qquad\quad\quad\cr
  ~\cr
  {-{4\sqrt{j(j+1)} \over 2j+1}\int dr\,\Phi_j^\dagger \left(\matrix{
  1&0\cr 0&0\cr }\right)\,V\Phi_{j+1}},\qquad\quad\quad\cr }
  \right.
\end{eqnarray}
and
\begin{eqnarray}
  \left<\Psi_{\ell\,'}^-\right|&&\left(\vec\alpha_q\cdot\vec n\right)
  \left(\vec\alpha_Q\cdot\vec n\right)V\left|\Psi_\ell^+\right> \nonumber \\
  ~\nonumber \\
  &&=\left\{\matrix{{i\over 2k+1}\int dr\,\Phi_{-k}^\dagger \left(\matrix{
  0&-1\cr 1&0\cr }\right)\,V\Phi_k \quad {\rm for}~
  k=-(j+1),\;j, \cr 
  ~\cr
  {2i\sqrt{j(j+1)} \over 2j+1}\int dr\,\Phi_{-j}^\dagger \left(\matrix{
  0&-1\cr 1&0\cr }\right)\,V\Phi_{-(j+1)},\qquad\quad\quad\cr}
  \right.
\end{eqnarray}
and their complex conjugates.
\subsubsection{First order terms }
\label{app:subsubsec:1st-+}
The first order $\Lambda_--\Lambda_+$ matrix element is given by
\begin{eqnarray}
  \left<{\Psi^-_{\ell\,'}}\right|H_1^{-+}\left|\Psi^+_\ell\right>
  &=&\left<{\Psi^-_{\ell\,'}}\right|{1 \over m_Q}
  \left[-S\beta_q\,\left(\vec\alpha_Q\cdot\vec p\right)
  +{i\over 2}\left(\beta_q\,S^\prime+V^\prime\right)
  \left(\vec\alpha_Q\cdot\vec n\right)\right]
  \left|\Psi^+_\ell\right>,\nonumber \\
  ~\nonumber \\
  &=&{1\over 2}{\rm tr}\int d^3r{\Psi _{j\,m}^{k^\prime}}^\dagger\,
  {1 \over m_Q}\left[- S\beta_q\,\left(\vec\sigma_Q \cdot\vec p\right)
  +{i\over 2}\left(\beta_q\,S'+V'\right)\left(\vec\sigma_Q\cdot\vec n\right)
  \right] \otimes\Psi _{j\,m}^k.
\end{eqnarray}
Non-vanishing matrix elements are given by, for $k=-(j+1)$, or $j$,
\begin{eqnarray}
  \left<{\Psi^-_{\ell\,'}}\right|&&
  S\beta_q\,\left(\vec\alpha_Q \cdot\vec p\right)\left|\Psi^+_\ell\right>
  \nonumber \\
  && \nonumber \\
  &&=\left\{\matrix{{-i\over 2k+1}\int dr\,\Phi_{-k}^\dagger S\left(\matrix{
  \partial_r+{k+1\over r}&0\cr 0&-\partial_r+{k-1\over r}\cr}\right)\Phi_k,\cr
  ~\cr
  {-2i\sqrt{j(j+1)}\over 2j+1}\int dr\,\Phi_{k+1}^\dagger S\left(\matrix{
  \partial_r-{k\over r}&0\cr 0&-\partial_r+{k-1\over r}\cr}\right)
  \Phi_k,\cr}\right.
\end{eqnarray}
and
\begin{eqnarray}
  \left<{\Psi^-_{\ell\,'}}\right|&&\left(\beta_q\,S'+V'\right)
  \left(\vec\alpha_Q \cdot\vec n\right)\left|\Psi^+_\ell\right>
  \nonumber \\
  && \nonumber \\
  &&=\left\{\matrix{{1\over 2k+1}\int dr\,\Phi_k^\dagger 
  \left(\matrix{S'+V'&0\cr 0&-S'+V'}\right)\Phi_k, \cr
  ~\cr
  {2\sqrt{j(j+1)}\over 2j+1}\int dr\,\Phi_j^\dagger 
  \left(\matrix{S'+V'&0\cr 0&-S'+V'}\right)
  \Phi_{-(j+1)},\quad\qquad\quad\cr}\right.
\end{eqnarray}
and their complex conjugates. Note that when one takes complex conjugate,
derivative operators do not operate on $S'$ and $V'$ but on a wave 
function, $\Phi_\ell$.

Matrix elements of interaction terms, $H_1^{--}$ and $H_2^{--}$,
between eigenfunctions ($\Psi_{j\;m}^k$) obtained above are summarized below.
Degeneracy can be resolved by diagonal as well as off-diagonal matrix 
elements of the last terms of
$H_1^{--}$ and $H_2^{--}$ together with contributions from negative components
of the wave functions coming from $H_0^{-+}$ and $H_1^{-+}$ 
as mentioned earlier. 
Here matrix elements computed by eigenfunctions with 
the same $k$'s are called ``diagonal'' and others with different $k$'s 
are ``off-diagonal''.
Calculating all the matrix elements from the hamiltonian given in the Section
\ref{sec:intro}, total mass matrix is given by, up to the second order 
in $1/m_Q$,
\begin{equation}
\left( {\matrix{{\scriptstyle E_{-1}+U_{-1,0}}
  &0&0&0&0&0&0&0\cr
  0&{\scriptstyle E_{-1}+U_{-1,1}}
  &0&0&0&0&{\scriptstyle V^1_{-1,2}}&0\cr
  0&0&{\scriptstyle E_1+U_{1,0}}&0&0&0&0&0\cr
  0&0&0&{\scriptstyle E_1+U_{1,1}}&{\scriptstyle V^1_{1,-2}}&0&0&0\cr
  0&0&0&{\scriptstyle V^1_{-2,1}}&{\scriptstyle E_{-2}+U_{-2,1}}&0&0&0\cr
  0&0&0&0&0&{\scriptstyle E_{-2}+U_{-2,2}}&0&0\cr
  0&{\scriptstyle V^1_{2,-1}}&0&0&0&0&{\scriptstyle E_2+U_{2,1}}&0\cr
  0&0&0&0&0&0&0&{\scriptstyle E_2+U_{2,2}}\cr }} \right),
\end{equation}
where
\begin{equation}
E_k=m_Q+E^k_0+E^k_1+E^k_2,
\quad U_{k,\;j}=U_{k,\;j}^{(1)}+U_{k,\;j}^{(2)},\quad
  V^j_{k,\;k'}=V_{k,\;k'}^{(1)\;j}+V_{k,\;k'}^{(2)\;j},
\end{equation}
Here superscripts mean the order of $1/m_Q$, $k$ and $k'$ stand for
$k$ quantum number, and $j$ for the total angular momentum.
For instance, $V_{k,\;k'}^{(2)\;j}$ means the matrix element of the third term
of $H_2^{--}$ between ${\Psi_{j\;m}^k}^\dagger$ and $\Psi_{j\;m}^{k'}$ given
by Eq.~(\ref{eq:app:2nd}). 

As for the $\Lambda_--\Lambda_+$ matrix element, 
we only need non-vanishing matrix elements of $H_0^{-+}$ or $c_{1-}^{\ell\,k}$
as one can see from Eqs.~(\ref{eq:app:c1-2} $\sim$ \ref{eq:app:E22})
up to the second order in $1/m_Q$, which are given below.
\begin{equation}
\left( {\matrix{
  0&0&{\scriptstyle c_{1-}^{-1,1}(0)}&0&0&0&0&0\cr
  0&0&0&{\scriptstyle c_{1-}^{-1,1}(1)}&
  {\scriptstyle c_{1-}^{-1,-2}(1)}&0&0&0\cr
  {\scriptstyle c_{1-}^{1,-1}(0)}&0&0&0&0&0&0&0\cr
  0&{\scriptstyle c_{1-}^{1,-1}(1)}&0&0&0&0&
  {\scriptstyle c_{1-}^{1,2}(1)}&0\cr
  0&{\scriptstyle c_{1-}^{-2,-1}(1)}&0&0&0&0&
  {\scriptstyle c_{1-}^{-2,2}(1)}&0\cr
  0&0&0&0&0&0&0&{\scriptstyle c_{1-}^{-2,2}(2)}\cr
  0&0&0&{\scriptstyle c_{1-}^{2,1}(1)}&
  {\scriptstyle c_{1-}^{2,-2}(1)}&0&0&0\cr
  0&0&0&0&0&{\scriptstyle c_{1-}^{2,-2}(2)}&0&0 }}\right),
\end{equation}
where the integer in the brackets is a value of a total angular momentum,
$j$, and it turns out that this matrix is hermitian, i.e., 
$c_{1-}^{\ell\,k}(j_0)=c_{1-}^{k\,\ell}(j_0)^*$ for the same value of $j=j_0$.

\begin{table}
\caption{Input values to determine parameters (Units are in GeV.)}
\label{table1}
\begin{tabular}{r r r r r r r}
  $m_q=m_u=m_d$ & $M_D$ & $M_{D^*}$ & 
  $M_{D_s}$ & $M_{D_s^*}$ & $M_B$ & $M_{B^*}$ \\ \hline
  0.01 & 1.867 & 2.008 & 1.969 & 2.112 & 5.279 & 5.325 \\
\end{tabular}
\end{table}

\begin{table}
\caption{Most optimal values of parameters determined by the least chi 
square method.}
\label{table2}
\begin{tabular}{l c c c c c c c}
  parameters & $\alpha_s$ &  $a$ (${\rm GeV}^{-1}$) &  $b$ (GeV) & 
  $m_c$ (GeV) & $m_s$ (GeV) & $m_b$ (GeV) & $\chi^2$ \\ \hline
  first order  & 0.3998 & 2.140 & -0.04798 & 1.457 & 
  0.09472 & 4.870 & $2.5\times 10^{-12}$ \\ \hline
  second order & 0.2834 & 1.974 & -0.07031 & 1.347 & 
  0.08988 & 4.753 & $6.8\times 10^{-9}$ \\
\end{tabular}
\end{table}

\begin{table}
\caption{$D$ meson mass spectrum (first order)}
\label{table3}
\begin{tabular}{r c c c r r}
  State ($J^P$) & $M_0$ & $p_1/M_0$ &
  $n_1/M_0$ & $M_{\rm calc}$ & $M_{\rm obs}$ \\ \hline
  $~^1S_0$ ($0^-$)  & 1.869 & -2.01$~(\times 10^{-2})$ &
                              1.93 $~(\times 10^{-2})$  & 1.867 & 1.867 \\
  $~^3S_1$ ($1^-$)  &       & 7.37 & 0.101 &2.008 & 2.008 \\
  $~^3P_0$ ($0^+$)  & 2.276 & -0.373 & 1.59 & 2.304 & - \\
  $"^3P_1"$ ($1^+$) &       & 7.50 & 0.0883 & 2.449 & 2.422(?) \\
  $"^1P_1"$ ($1^+$) & 2.216 & 7.51 & 0.0210 & 2.383 & - \\
  $~^3P_2$ ($2^+$)  &       & 9.54 & 4.72$\times 10^{-7}$ & 2.428 & 2.459 \\
  $~^3D_1$ ($1^-$)  & 2.440 & 181 & 0.0242 & 6.850 & - \\
  $"^3D_2"$ ($2^-$) &       & 177 & 4.28$\times 10^{-7}$ & 6.747 & - \\
\end{tabular}
\end{table}

\begin{table}
\caption{$D_s$ meson mass spectrum (first order)}
\label{table4}
\begin{tabular}{r c c c r r}
  state ($J^P$) & $M_0$ & $p_1/M_0$ & 
  $n_1/M_0$ & $M_{\rm calc}$ & $M_{\rm obs}$ \\ \hline
  $~^1S_0$ ($0^-$)  & 1.986 & -2.66$~(\times 10^{-2})$ &
                               1.67$~(\times 10^{-2})$  & 1.966 & 1.969 \\
  $~^3S_1$ ($1^-$)  &       & 6.95 & 0.0851 & 2.125 & 2.112(?) \\
  $~^3P_0$ ($0^+$)  & 2.288 & 0.769 & 1.45 & 2.339 & - \\
  $"^3P_1"$ ($1^+$) &       & 8.62 & 0.0835 & 2.487 & 2.535 \\
  $"^1P_1"$ ($1^+$) & 2.335 & 6.87 & 0.0194 & 2.496 & - \\
  $~^3P_2$ ($2^+$)  &       & 8.79 & 8.87$\times 10^{-7}$ & 2.540 
  & 2.574(?) \\
  $~^3D_1$ ($1^-$)  & 2.401 & 159 & 0.0282 & 6.230 & - \\
  $"^3D_2"$ ($2^-$) &       & 126 & 8.63$\times 10^{-7}$ & 5.428 & - \\
\end{tabular}
\end{table}

\begin{table}
\caption{$B$ meson mass spectrum (first order)}
\label{table5}
\begin{tabular}{r c c c r r}
  state ($J^P$) & $M_0$ & $p_1/M_0$ & 
  $n_1/M_0$ & $M_{\rm calc}$ & $M_{\rm obs}$ \\ \hline
  $~^1S_0$ ($0^-$)  & 5.281 & -2.13$~(\times 10^{-3})$ &
                               2.05$~(\times 10^{-3})$ & 5.281 & 5.279 \\
  $~^3S_1$ ($1^-$)  &       &  7.80 & 0.107 & 5.323 & 5.325 \\
  $~^3P_0$ ($0^+$)  & 5.689 & -0.447 & 1.90 & 5.697 & - \\
  $"^3P_1"$ ($1^+$) &       &  8.98 & 0.106 & 5.740 & - \\
  $"^1P_1"$ ($1^+$) & 5.629 &  8.85 & 0.0247 & 5.679 & - \\
  $~^3P_2$ ($2^+$)  &       &  11.2 & 5.56$\times 10^{-8}$ & 5.692 & - \\
  $~^3D_1$ ($1^-$)  & 5.853 &  225 & 0.0302 & 7.172 & - \\
  $"^3D_2"$ ($2^-$) &       &  220 & 5.34$\times 10^{-8}$ & 7.141 & - \\
\end{tabular}
\end{table}

\begin{table}
\caption{$B_s$ meson mass spectrum (first order)}
\label{table6}
\begin{tabular}{r c c c r r}
  state ($J^P$) & $M_0$ & $p_1/M_0$ & 
  $n_1/M_0$ & $M_{\rm calc}$ & $M_{\rm obs}$ \\ \hline
  $~^1S_0$ ($0^-$)  & 5.399 & -2.92$~(\times 10^{-3})$ &
                               1.84$~(\times 10^{-3})$ & 5.393 & 5.369 \\
  $~^3S_1$ ($1^-$)  &       & 7.65 & 0.0936 & 5.440 & - \\
  $~^3P_0$ ($0^+$)  & 5.701 & 0.923 & 1.74 & 5.716 & - \\
  $"^3P_1"$ ($1^+$) &       & 10.3  & 0.100 & 5.760 & - \\
  $"^1P_1"$ ($1^+$) & 5.748 & 8.35  & 0.0236 & 5.796 & - \\
  $~^3P_2$ ($2^+$)  &       & 10.7  & 1.08$\times 10^{-7}$ & 5.809 & - \\
  $~^3D_1$ ($1^-$)  & 5.814 & 197  & 0.0348 & 6.959 & - \\
  $"^3D_2"$ ($2^-$) &       & 156  & 1.07$\times 10^{-7}$ & 6.719 & - \\
\end{tabular}
\end{table}

\begin{table}
\caption{$D$ meson mass spectrum (second order)}
\label{table7}
\begin{tabular}{r c c c c c r r}
  state ($J^P$) & $M_0$ & $p_1/M_0$ & $n_1/M_0$ & 
  $p_2/M_0$ & $n_2/M_0$ & $M_{\rm calc}$ & $M_{\rm obs}$ \\ \hline
  $~^1S_0$ ($0^-$)  & 1.865 & 0.457$~(\times 10^{-4})$ &
  1.03$~(\times 10^{-2})$ &  -1.86$(\times 10^{-17})$ & 
  -1.40$~(\times 10^{-2})$ & 1.867 & 1.867 \\
  $~^3S_1$ ($1^-$) & & 7.41 & 0.0540 & -2.95$\times 10^{-4}$ & 0.209 & 2.008 
  & 2.008 \\
  $~^3P_0$ ($0^+$) & 2.319 & 2.33 & 0.828 & 3.44 & 0.689 & 2.408 & - \\
  $"^3P_1"$ ($1^+$) & & 7.00 & 0.164 & 3.63$\times 10^{-4}$ & -0.0107 & 2.486
  & 2.422(?) \\
  $"^1P_1"$ ($1^+$) & 2.205 & 8.22 & 0.0532 & -3.82$\times 10^{-4}$ & -0.0541
  & 2.388 & - \\
  $~^3P_2$ ($2^+$)  &       & 8.67 & 0.0156 & -0.157 & 0.112 & 2.399 
  & 2.459 \\
  $~^3D_1$ ($1^-$)  & 2.584 & 8.29 & 0.154 & 2.14$\times 10^{-4}$ & -0.177 
  & 2.798 & - \\
  $"^3D_2"$ ($2^-$) &       & 8.13 & 0.0133 & 3.17 & -0.134 & 2.791 & - \\
\end{tabular}
\end{table}

\begin{table}
\caption{$D_s$ meson mass spectrum (second order)}
\label{table8}
\begin{tabular}{r c c c c c r r}
  State ($J^P$) & $M_0$ & $p_1/M_0$ & $n_1/M_0$ & 
  $p_2/M_0$ & $n_2/M_0$ & $M_{\rm calc}$ & $M_{\rm obs}$ \\ \hline
  $~^1S_0$ ($0^-$)  & 1.976 & 0.150$~(\times 10^{-2})$ &
  90.6$~(\times 10^{-4})$ & 2.81$~(\times 10^{-17})$ & 
  -1.60$~(\times 10^{-2})$ & 1.965 & 1.969 \\
  $~^3S_1$ ($1^-$)  &       &  7.76 & 4.65 & -7.02$\times 10^{-4}$ & 0.243 
  & 2.133 & 2.112(?) \\
  $~^3P_0$ ($0^+$)  & 2.331 &  3.20 & 76.8 & 2.08 & 0.956 & 2.446 & - \\
  $"^3P_1"$ ($1^+$) &       &  8.12 & 3.87 & 2.20$\times 10^{-3}$ & 0.0361 
  & 2.527 & 2.535 \\
  $"^1P_1"$ ($1^+$) & 2.317 &  7.29 & 0.925 & -2.22$\times 10^{-3}$ & 0.0229 
  & 2.482 & - \\
  $~^3P_2$ ($2^+$)  &       &  8.31 & 2.96$\times 10^{-8}$ & -5.98 & -0.00166
  & 2.509 & 2.574(?) \\
  $~^3D_1$ ($1^-$)  & 2.582 &  702  & 0.770 & 5.38$\times 10^{-4}$ & -0.00632
  & 2.072 & - \\
  $"^3D_2"$ ($2^-$) &       & -4470 & 2.65$\times 10^{-8}$ & 212 & 0.000139 
  & -113 & - \\
\end{tabular}
\end{table}

\begin{table}
\caption{$B$ meson mass spectrum (second order)}
\label{table9}
\begin{tabular}{r c c c c c r r}
  State ($J^P$) & $M_0$ & $p_1/M_0$ & $n_1/M_0$ & 
  $p_2/M_0$ & $n_2/M_0$ & $M_{\rm calc}$ & $M_{\rm obs}$ \\ \hline
  $~^1S_0$ ($0^-$)  & 5.271 & 0.212$~(\times 10^{-2})$ &
  10.3$~(\times 10^{-4})$ & 2.80$~(\times 10^{-17})$ & 
  -3.97$~(\times 10^{-4})$ & 5.286 & 5.279 \\
  $~^3S_1$ ($1^-$)  &       &  0.862 & 0.541 & -8.41$\times 10^{-6}$ & 0.593 
  & 5.317 & 5.325 \\
  $~^3P_0$ ($0^+$)  & 5.725 &  0.387 & 9.50 & -3.54 & 2.24 & 5.754 & - \\
  $"^3P_1"$ ($1^+$) &       &  0.965 & 1.89 & 1.18$\times 10^{-5}$ & -0.0347 
  & 5.782 & - \\
  $"^1P_1"$ ($1^+$) & 5.611 &  1.08  & 0.593 & -1.21$\times 10^{-5}$ & 0.171 
  & 5.672 & - \\
  $~^3P_2$ ($2^+$)  &       &  1.23  & 0.174 & -1.06 & 0.352 & 5.680 & - \\
  $~^3D_1$ ($1^-$)  & 5.990 &  1.18  & 1.88  & 7.40$\times 10^{-6}$ & -0.614 
  & 6.061 & - \\
  $"^3D_2"$ ($2^-$) &       &  1.28  & 0.163 & -6.48 & -0.465 & 6.066 & - \\
\end{tabular}
\end{table}

\begin{table}
\caption{$B_s$ meson mass spectrum (second order)}
\label{table10}
\begin{tabular}{r c c c c c r r}
  State ($J^P$) & $M_0$ & $p_1/M_0$ & $n_1/M_0$ & 
  $p_2/M_0$ & $n_2/M_0$ & $M_{\rm calc}$ & $M_{\rm obs}$ \\ \hline
  $~^1S_0$ ($0^-$)  & 5.382 & 0.161$~(\times 10^{-2})$ &
  9.43$~(\times 10^{-4})$ & -7.46$~(\times 10^{-17})$ & 
  -4.72$~(\times 10^{-4})$ & 5.393 & 5.369 \\
  $~^3S_1$ ($1^-$)  &       & 0.882 & 0.483 & -2.07$\times 10^{-5}$ & 0.717 
  & 5.430 & - \\
  $~^3P_0$ ($0^+$)  & 5.739 & 0.505 & 8.84 & 3.12$\times 10^{-4}$ & 3.12 
  & 5.773 & - \\
  $"^3P_1"$ ($1^+$) &       & 1.11  & 0.446 & 7.19$\times 10^{-5}$ & 0.118 
  & 5.801 & - \\
  $"^1P_1"$ ($1^+$) & 5.723 & 1.03 & 0.106 & -7.21$\times 10^{-5}$ & 0.0743 
  & 5.782 & - \\
  $~^3P_2$ ($2^+$)  &       & 1.19 & 3.39$\times 10^{-9}$ & -3.78 & -0.00540 
  & 5.791 & - \\
  $~^3D_1$ ($1^-$)  & 5.988 & 43.9 & 0.0940 & 1.86$\times 10^{-5}$ & -0.0219 
  & 8.618 & - \\
  $"^3D_2"$ ($2^-$) &       & -136 & 3.24$\times 10^{-9}$ & -2.06 
  & 4.82$\times 10^{-8}$ & -2.174 & - \\
\end{tabular}
\end{table}

\begin{table}
\caption{States classified by various quantum numbers.}
\label{table0}
\begin{tabular}{|c|c|c|c|c|c|c|c|c|}
  $k$           &    -1    &    -1     &     1     &     1    &
      -2      &    -2     &     2     &     2    \\ \hline
  $j$           &     0    &     1     &     0     &     1    &
       1      &     2     &     1     &     2    \\ \hline
  $J^P$         & $0^-$    & $1^-$     & $0^+$     & $1^+$    &
   $1^+$      & $2^+$     & $1^-$     & $2^-$    \\ \hline
  $~^{2S+1}L_J$ & $~^1S_0$ & $~^3S_1$  & $~^3P_0$  & $~^3P_1$, $~^1P_1$ &
   $~^1P_1$, $~^3P_1$  & $~^3P_2$  & $~^3D_1$  & $~^3D_2$, $~^1D_2$ \\
\end{tabular}
\end{table}


\begin{references}
\bibitem{Morii3}
    Eq.~(\ref{eq:app:ham+-:h1--}) of our paper corresponds to Eq.~(3d) 
    in the paper by Morishita et. al. \cite{Morii1} which includes errors and
    we corrected it by adding the last terms of the curly brackets.
\end{references}
\end{document}